\documentclass[aps,prb,twocolumn,showpacs,citeautoscript,reprint,longbibliography,superscriptaddress]{revtex4-1}
\usepackage{graphicx}
\usepackage{bm,amsmath,amssymb,wasysym,amsfonts,dcolumn}
\usepackage[colorlinks=true, urlcolor=blue, linkcolor=blue, citecolor=blue, pdfborder={0 0 0}]{hyperref}
\usepackage[usenames,dvipsnames]{xcolor}
\usepackage{cleveref}
\usepackage{dcolumn}
\newcolumntype{d}[1]{D..{#1}}
\crefname{figure}{Fig.}{Figs}
\crefname{table}{Table}{Tables}

\newcommand{\fat}[1]{\mathbf{#1}}

\begin{document}

\title{Calculating spin transport properties from first principles: spin currents}

\author{R.J.H. Wesselink}
\altaffiliation{These authors contributed equally to this work.}
\affiliation{Faculty of Science and Technology and MESA$^+$ Institute for Nanotechnology, University of Twente, P.O. Box 217,
		7500 AE Enschede, The Netherlands}

\author{K. Gupta}
\altaffiliation{These authors contributed equally to this work.}
\affiliation{Faculty of Science and Technology and MESA$^+$ Institute for Nanotechnology, University of Twente, P.O. Box 217,
		7500 AE Enschede, The Netherlands}

\author{Z. Yuan}
\affiliation{Faculty of Science and Technology and MESA$^+$ Institute for Nanotechnology, University of Twente, P.O. Box 217,
		7500 AE Enschede, The Netherlands}
\affiliation{The Center for Advanced Quantum Studies and Department of Physics, Beijing Normal University, 100875 Beijing, China}


\author{Paul J. Kelly}
\affiliation{Faculty of Science and Technology and MESA$^+$ Institute for Nanotechnology, University of Twente, P.O. Box 217,
		7500 AE Enschede, The Netherlands}
\affiliation{The Center for Advanced Quantum Studies and Department of Physics, Beijing Normal University, 100875 Beijing, China}

\date{\today}

\begin{abstract}
Local charge and spin currents are evaluated from the solutions of fully relativistic quantum mechanical scattering calculations for systems that include temperature-induced lattice and spin disorder as well as intrinsic alloy disorder. This makes it possible to determine material-specific spin transport parameters at finite temperatures. Illustrations are given for a number of important materials and parameters at 300 K. 
The spin-flip diffusion length $l_{\rm sf}$ of Pt is determined from the exponential decay of a spin current injected into a long length of thermally disordered Pt; we find $l_{\rm sf}^{\rm Pt}= 5.3\pm0.4 \,$nm. 
For the ferromagnetic substitutional disordered alloy Permalloy (Py), we inject currents that are fully polarized parallel and antiparallel to the magnetization and calculate $l_{\rm sf}$ from the exponential decay of their difference; we find $l_{\rm sf}^{\rm Py}= 2.8 \pm 0.1 \,$nm. 
The transport polarization $\beta$ is found from the asymptotic polarization of a charge current in  a long length of Py to be $\beta = 0.75 \pm 0.01$. 
The spin Hall angle $\Theta_{\rm sH}$ is determined from the transverse spin current induced by the passage of a longitudinal charge current in thermally disordered Pt; our best estimate is $\Theta_{\rm sH}^{\rm Pt}=4.5 \pm 1 \%$ corresponding to the experimental room temperature bulk resistivity $\rho =10.8 \mu \Omega \,$cm. 
\end{abstract}

\pacs{}

\maketitle

\section{Introduction} 
\label{Sec:Intro}

Experiments in the field of spintronics are almost universally interpreted using semiclassical transport theories \cite{Brataas:prp06}. In such phenomenological theories based upon the Boltzmann or diffusion equations, a number of parameters are used to describe how transport depends on material composition, structure and temperature. For a bulk nonmagnetic material (NM) these are the resistivity $\rho$, the spin flip diffusion length (SDL) $l_{\rm sf}$ \cite{vanSon:prl87, Valet:prb93, Bass:jpcm07} and the spin Hall angle (SHA) $\Theta_{\rm sH}$ that measures the efficiency of the spin Hall effect (SHE) \cite{Dyakonov:pla71, Hirsch:prl99, Zhang:prl00} whereby a longitudinal charge current is converted to a transverse spin current, or of its inverse \cite{Hoffmann:ieeem13, Sinova:rmp15}. The transport properties of a ferromagnetic material (FM) are characterized in terms of the spin-dependent resistivities $\rho_{\downarrow}$ and $\rho_{\uparrow}$, a SDL $l_{\rm sf}$ and an anomalous Hall angle (AHA). Instead of $\rho_{\downarrow}$ and $\rho_{\uparrow}$, the polarization $\beta = (\rho_{\downarrow} - \rho_{\uparrow})/(\rho_{\downarrow} + \rho_{\uparrow})$ and a resistivity $\rho^* = (\rho_{\uparrow} + \rho_{\downarrow})/4$ are frequently used.
Phenomenological theories ultimately aim to relate currents of charge ${\bf j}_c$ and spin ${\bf j}_{s \alpha}$ to, respectively, gradients of the chemical potential $\mu_c$ and spin accumulation $\mu_{s \alpha}$ (where $\alpha$ labels the spin component) in terms of the above parameters but they tell us nothing about the values of the parameters for particular materials or combinations of materials. This paper is concerned with evaluating these parameters using realistic electronic structures and models of disorder within the framework of density functional theory (DFT).

Ten years ago only a handful of measurements had been made of $l_{\rm sf}$, $\Theta_{\rm sH}$ and $\beta$ and a wide range of values was found for all three parameters. The polarization was found to depend on the type of measurement used to extract it and this usually involved an interface \cite{Mazin:prl99}. The introduction of current-induced spin-wave Doppler shift measurements \cite{Vlaminck:sc08} made it possible to probe the current polarization in the bulk of a magnetic material far from any interfaces. The advent of nonlocal spin injection and spin-pumping (SP) allowed the SHA and SDL to be studied by means of the inverse SHE (ISHE). Alternatively, spin currents generated by the SHE could be used to drive the precession of a magnetization by the spin-transfer torque (STT). These innovations have changed the situation radically over the past ten years yielding a host of new, mainly room temperature (RT) results \cite{Haidar:prb13, Hoffmann:ieeem13, Sinova:rmp15}. All of these methods involve NM$|$FM interfaces that introduce a variety of interface-related factors such as spin memory loss and interface spin Hall effects that are not taken into account systematically in the interpretation of the experimental results leading to a large spread in estimates of the SDL and SHA \cite{Rojas-Sanchez:prl14}. Perhaps as a result of this, there are few systematic studies of the temperature dependence of  $l_{\rm sf}$, $\Theta_{\rm sH}$ and $\beta$ \cite{Isasa:prb15a, *Isasa:prb15b}.     
   
To simultaneously describe the magnetic and transport properties of transition metals quantitatively requires taking into account their degenerate electronic structures and complex Fermi surfaces. Realistic electronic structures have only been incorporated into Boltzmann transport theory for the particular cases of point impurities \cite{Mertig:rpp99} and for thermally disordered elemental metals \cite{Savrasov:prb96b}. For the layered structures that form the backbone of spintronics, the most promising way to combine complex electronic structures with transport theory is to use scattering theory formulated either in terms of nonequilibrium Green's functions or wave-function matching \cite{Brataas:prp06} that are equivalent in the linear response regime \cite{Khomyakov:prb05}. The effect of temperature on transport has been successfully included in scattering calculations in the adiabatic approximation by constructing scattering regions with temperature-induced lattice and spin disorder \cite{LiuY:prb11, LiuY:prb15}. By constructing charge and spin currents (and chemical potentials \cite{YuanZ:tbp19}) from the scattering theory solutions, we aim to make contact with the phenomenological theories that are formulated in terms of these quantities. Though we will be focusing on bulk transport properties in this manuscript, the methodology we present can be directly extended to interfaces \cite{WangL:prl16, Gupta:tbp19}.

 Indeed, in a two-terminal $\mathcal{L}|\mathcal{S}|\mathcal{R}$ scattering formalism where a ``scattering'' region $\mathcal{S}$ is probed by attaching left ($\mathcal{L}$) and right ($\mathcal{R}$) leads to study how incoming Bloch states in the leads are scattered into outgoing states, interfaces are unavoidable and must be factored into (or out of) any subsequent analysis. For example, an interface gives rise to an interface resistance even in the absence of disorder because of the electronic structure mismatch between different materials \cite{Schep:prb97, Xia:prb01, Xia:prb06, Xu:prl06}. For disordered materials, the linear dependence of the resistance $R$ on the length $L$ of the scattering region allows the interface contribution to be factored out by extracting the bulk resistivity from the linear part of $R(L)$ \cite{Starikov:prl10, Starikov:prb18}. An analogous procedure can be applied to study the magnetization damping \cite{Starikov:prl10, LiuY:prl14, Starikov:prb18} where interfaces give rise to important observable effects \cite{Brataas:prp06}. 
  
\begin{figure}[t]
\begin{center}
\includegraphics[width=8.5 cm]{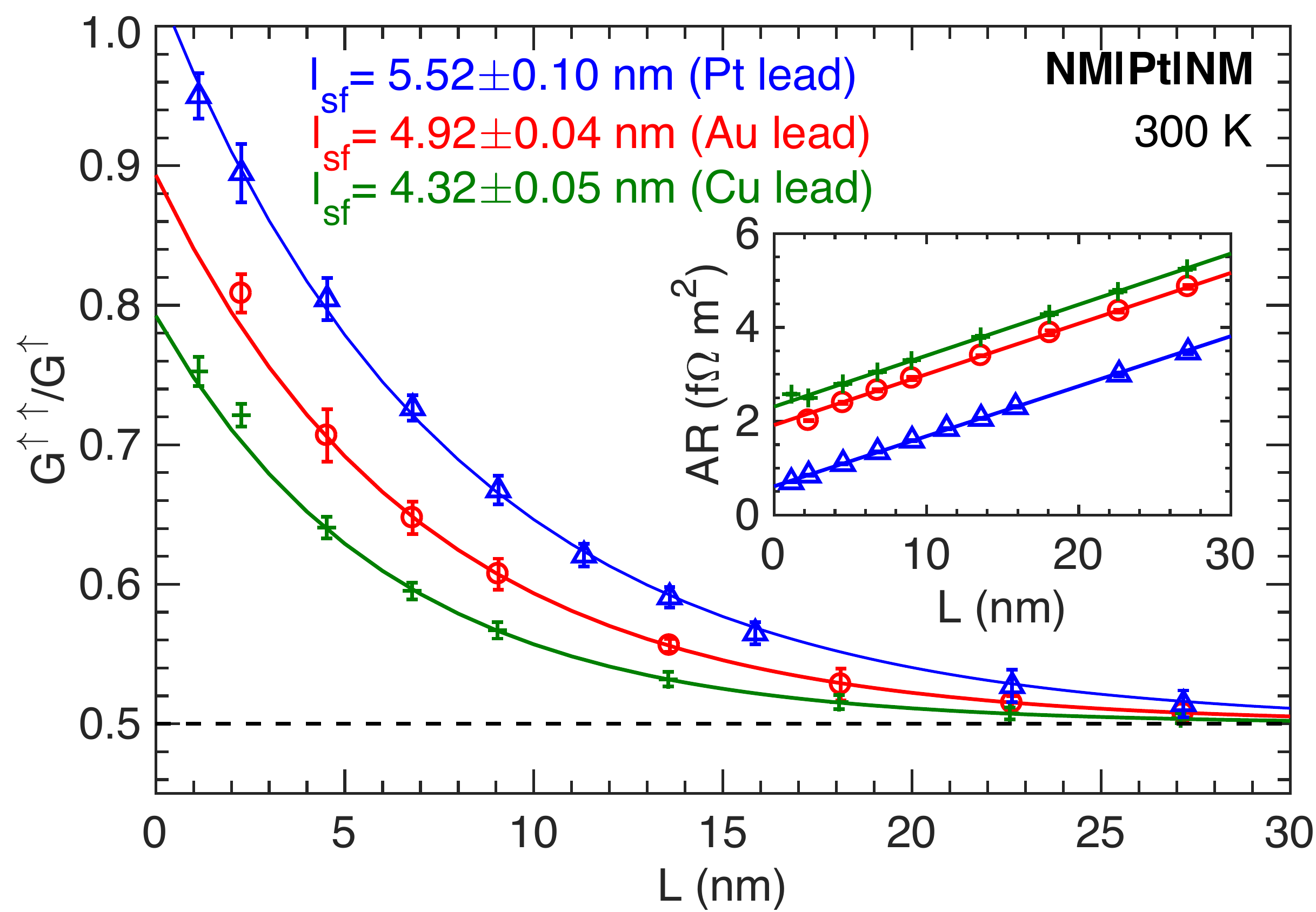} 
\end{center}
\caption{Calculated fractional spin conductance $G^{\uparrow\uparrow}/G^{\uparrow}$ for RT  thermally disordered Pt sandwiched between the different ballistic leads: Pt (blue triangles), Au (red circles) and Cu (green crosses). $G^{\sigma\sigma'}$ is ($e^2/h$ times) the transmission probability of a spin $\sigma$ from the left hand lead into a spin $\sigma'$ in the right hand lead; $G^{\uparrow}=G^{\uparrow\uparrow}+G^{\uparrow\downarrow}$. The solid lines are the exponential fits to the calculated values giving rise to $l_{\rm sf}$. Inset: The areal resistance of the NM$|$Pt$|$NM as a function of the length $L$ of Pt for all three ballistic leads. Solid lines are linear fits whose slopes yield identical resistivities in the three cases. Data for $L<4$ nm is excluded from the linear fit \cite{Starikov:prb18}.
}
\label{Fig1}
\end{figure}

 In the case of spin-flipping, the exponential dependence on $L$ of the transmission probability $T^{\sigma \sigma'}$ of states with spin $\sigma$ from one lead into states with spin $\sigma'$ in the other lead  makes this numerically challenging. Starikov, Liu and co-workers used $T^{\sigma \sigma'}$ to evaluate the SDL in Fe$_x$Ni$_{1-x}$ disordered alloys \cite{Starikov:prl10} and in thermally disordered Pd and Pt \cite{LiuY:prl14}. In terms of the corresponding spin-resolved conductances $G^{\sigma \sigma'}=\frac{e^2}{h}T^{\sigma \sigma'}$, the total conductance of spin $\sigma$ is given by $G^{\sigma}=\sum_{\sigma'}G^{\sigma \sigma'}$ and the total conductance of the system is the sum over both possible spins: $G=\sum_{\sigma\sigma'}G^{\sigma \sigma'}$. For a single spin channel, Liu et al. identified the exponential decay of the ``fractional spin conductance'' $G^{\uparrow\uparrow}/G^{\uparrow}$ with the ``spin diffusion length'' $l_\uparrow$. In Fig.~\ref{Fig1} we show $G^{\uparrow\uparrow}/G^{\uparrow}$ for RT thermally disordered Pt and different lead materials. The lattice disorder in the scattering region is taken to be Gaussian with a mean-square displacment chosen to reproduce the experimental room temperature resistivity $\rho=10.8~\mu \Omega$~cm \cite{HCP90}. Using ballistic Pt leads, we calculate the (blue) curve indicated with open triangles in Fig.~\ref{Fig1} from which we obtain a value of $l_\uparrow = 7.8 \pm 0.3$~nm. Because Pt is spin degenerate, $l_\downarrow \equiv l_\uparrow$ and \cite{Valet:prb93} $l_{\rm sf}=(l_{\uparrow}^{-2}+l_{\downarrow}^{-2})^{-1/2}=5.52\pm0.10$~nm in agreement with Ref.~\onlinecite{LiuY:prl14}. For $L \sim 1 \,$nm, we see that $G^{\uparrow\uparrow} \sim G^{\uparrow}$ indicative of a very weak interface between ballistic Pt leads and thermally disordered Pt.
When we use Au leads however, the effect of the interface becomes more noticeable and the value of $l_{\rm sf}$ is reduced to $\sim 4.9\,$nm. Because of the large difference of $\sim 8.5\%$ between the lattice constants of Pt and Cu, to study an interface between them we use an $8 \times 8$ lateral supercell of Cu to match to a $2\sqrt{13} \times 2\sqrt{13}$ lateral supercell of Pt. In this case, the interface is even stronger and we find an even shorter value of $l_{\rm sf}\sim 4.3\,$nm. This dependence of $l_{\rm sf}$ on the lead material is unsatisfactory. 

For an ohmic material, the conductance decays as $1/L$ and it is relatively easy to separate out interface effects by plotting the resistance $R = 1/G$ as a function of $L$ to determine the resistivity $\rho$, eventually ignoring short values of $L$ not characteristic of the bulk material as illustrated in the inset to Fig.~\ref{Fig1}. However, in the SDL case where the partial conductances decay exponentially, it is numerically much less straightforward to eliminate interface contributions. Ignoring too many small values of $L$ leaves us with too few data points with which to determine $l_{\rm sf}$ accurately. Unfortunately, we do not know a priori how far the effect of the interface extends. Similar considerations apply to the determination of $l_{\rm sf}$ for a ferromagnetic material when we examine the effect of using different lead materials. 
 
Local spin currents provide a description of the scattering region layer by layer. Contributions from interfaces show up only in layers close to the interfaces and not deep in the bulk. In the present paper we will resolve the problems discussed above by evaluating spin currents as a function of $z$ from the results of scattering calculations that include temperature-induced lattice and spin disorder as well as alloy disorder but do not assume diffusive behaviour a priori; in a companion paper, we will evaluate local chemical potentials in an analogous manner \cite{YuanZ:tbp19}. By focussing on the currents and chemical potentials employed in semiclassical theories \cite{vanSon:prl87} such as the Valet-Fert (VF) formalism \cite{Valet:prb93} that are widely used to interpret experiments, we will be able to evaluate the parameters that occur in those formalisms. For example, we will be able to determine the SDL $l_{\rm sf}$ from the exponential decay of a spin current injected into a long length of thermally disordered material. The transport polarization $\beta$ of the ferromagnetic substitutional disordered alloy Permalloy (Py, Fe$_{20}$Ni$_{80}$) will be determined straighforwardly from the asymptotic polarization of a charge current. The spin Hall angle $\Theta_{\rm sH}$ of Pt will be found from the transverse spin current induced by the passage of a longitudinal charge current. We will demonstrate that we can treat sufficiently long scattering regions as to be able to distinguish bulk and interface behaviour in practice. In a separate publication we will study the interface contributions explicitly in order to extract interface parameters for various FM$|$NM and NM$|$NM$'$ interfaces \cite{Gupta:tbp19}.

The plan of this paper is as follows. We begin Sec.~\ref{Sec:Methods} with a summary of the phenomenological Valet-Fert formalism (Sec.~\ref{SSec:theory}) containing the parameters we aim to evaluate. Sec.~\ref{SSec:scatt} outlines the quantum mechanical formalism that results in scattering wavefunctions which we will use to calculate position resolved charge and spin currents. In Sec.~\ref{SSec:curr} we explain how  currents between pairs of atoms are calculated using the scattering wavefunctions. Sec.~\ref{SSec:prac} explains how layer averaged currents are constructed from the interatomic currents. The most important practical aspects of scattering  calculations that determine the accuracy of the computational results are reviewed in Sec.~\ref{SSec:scatcalc}. In Sec.~\ref{Sec:calc} we illustrate the foregoing methodology by calculating $l_{\rm sf}$ for Pt (\ref{SSec:lsf_Pt}), $l_{\rm sf}$ (\ref{SSec:lsf_Py}) and $\beta$ (\ref{SSec:beta_Py}) for Py, and $\Theta_{\rm sH}$ for Pt (\ref{SSec:sha_Pt}). The emphasis in this paper will be on studying how the parameters depend on computational details of the scattering calculations such as lateral supercell size, Brillouin zone (BZ) sampling, basis set etc. A comparison with experiment and other calculations is made in Section~\ref{Sec:Comp}. Our results are summarized and some conclusions are drawn in Section~\ref{Sec:S&C}. 

\section{Methods}
\label{Sec:Methods}
 
\subsection{Semiclassical transport theory} 
\label{SSec:theory}

In this section, we recapitulate the VF description of spin transport that characterizes transport in terms of material-specific parameters. Starting from the Boltzmann formalism, Valet and Fert \cite{Valet:prb93} derived the following macroscopic equations for a current flowing along the $z$ direction perpendicular to the interface plane in an axially symmetric ``current perpendicular to the plane'' (CPP) geometry, 
\begin{subequations}
\begin{align}
\frac{\partial^2 \mu_s}{\partial z^2}&=\frac{\mu_s}{l_{\rm sf}^2}, 
\label{eq:diffusion}\\
j_{\sigma}(z)&=-\frac{1}{e \rho_{\sigma}}\frac{\partial\mu_{\sigma}}{\partial z}. 
\label{eq:ohm}
\end{align}
\end{subequations}
With respect to a quantization axis taken to be the $z$ axis, the majority and minority spin-polarized current densities and chemical potentials are denoted by $j_{\sigma}$ and $\mu_{\sigma}$ respectively with $\sigma = \uparrow$ (majority) or $\downarrow$ (minority). $\mu_s \equiv \mu_{sz}=\mu_\uparrow-\mu_\downarrow$ and $\rho_{\sigma}$ is the spin-dependent bulk resistivity.  According to the two-current series resistor model \cite{Valet:prb93}, resistances are first calculated separately for spin up and spin down electrons and then added in parallel. For non-magnetic materials, $\rho_{\uparrow}= \rho_{\downarrow} = 2\rho$, where $\rho$ is the total resistivity. Thus, spin transport in the bulk of a  material can be characterized in terms of its resistivity $\rho$ and SDL $l_{\rm sf}$. Equations (\ref{eq:diffusion}) and (\ref{eq:ohm}) can be solved for $\mu_\uparrow$, $\mu_\downarrow$, $j_\uparrow$, and $j_\downarrow$ making use of the condition that the total current density $j=j_\uparrow+j_\downarrow$ is conserved in one-dimensional transport. Dropping the ``sf'' subscript when there is no danger of confusion, the general solution of \eqref{eq:diffusion} is $\mu_s(z)=A e^{z/l}+B e^{-z/l}$. The normalized effective spin-current density $\widehat{j_s} \equiv j_{sz}^z/j= [j_\uparrow(z)-j_\downarrow(z)]/j$ is given by 
\begin{equation}
\widehat{j_s}(z) = \beta - \frac{1}{2ej\rho^* l}\Big[A e^{z/l} - B e^{-z/l}\Big]
\label{eq:js}
\end{equation}
where the coefficients $A$ and $B$ can be determined by using appropriate boundary conditions. For a NM material $\beta=0$. We will be concerned with calculating $j_s(z)$ from the results of two-terminal scattering calculations for $\mathcal{L}|\mathcal{S}|\mathcal{R}$ configurations. The coefficients $A_{\mathcal L}, B_{\mathcal L}, A_{\mathcal S}, B_{\mathcal S}, A_{\mathcal R}$ and $B_{\mathcal R}$ will be determined by imposing suitable boundary conditions at the $\mathcal{L}|\mathcal{S}$ and $\mathcal{S}|\mathcal{R}$ interfaces.  

\subsubsection*{Spin-flip diffusion length}
Equation (\ref{eq:js}) provides a simple prescription for extracting the SDL from a calculation of the spin current $\widehat{j_s}(z)$. In the case of a non-magnetic material, we choose the left lead to be ferromagnetic, e.g. a half metallic ferromagnet, so the current entering the non-magnetic material is fully polarized with $(|\widehat{j_s}(0)|=1)$. The right lead is nonmagnetic so $\widehat{j_s}(z) \rightarrow 0$ in the limit of large $z$. The boundary condition for the right lead in this limit is $\widehat{j_s}(\infty)=0$ so $\widehat{j_s}(z)=C \exp(-z/l)$ and $l_{\rm sf}$ can be determined from the slope of $\ln \widehat{j_s}(z)$. 

\subsubsection*{Polarization}

For a symmetric NM$|$FM$|$NM configuration with a thickness $L$ of FM, we choose the origin at the middle of the FM layer so $B=-A$ in \eqref{eq:js} and the spin current has the form 
\begin{equation}
\widehat{j_s}(z) = \frac{j_\uparrow(z)-j_\downarrow(z)}{j} = \beta - c \cosh \frac{z}{l}
\label{eq:jss}
\end{equation}
and $\widehat{j_s}(z) \rightarrow \beta$ for scattering regions much longer than $l_{\rm sf}$.

\subsubsection*{Spin-Hall angle}

The spin Hall effect is such that passage of a charge current through an NM$|$NM$'|$NM configuration leads to the generation of transverse spin currents $j_{s\alpha}^{\perp}$ where $\alpha$ labels the direction of spin polarization that is given by the vector product of the driving charge current (assumed to be in the $z$ direction) and the induced transverse spin current ($\perp$). For a constant charge current density $j$, the normalized transverse spin current sufficiently far from the interfaces gives the spin Hall angle $\Theta_{\rm sH}=\widehat{j}_s^{\perp}\equiv j_{s\alpha}^{\perp}/j$.  

\subsection{Quantum Mechanical Scattering}
\label{SSec:scatt}

\begin{figure}[b]
\begin{center}
\includegraphics[width = 8.2cm]{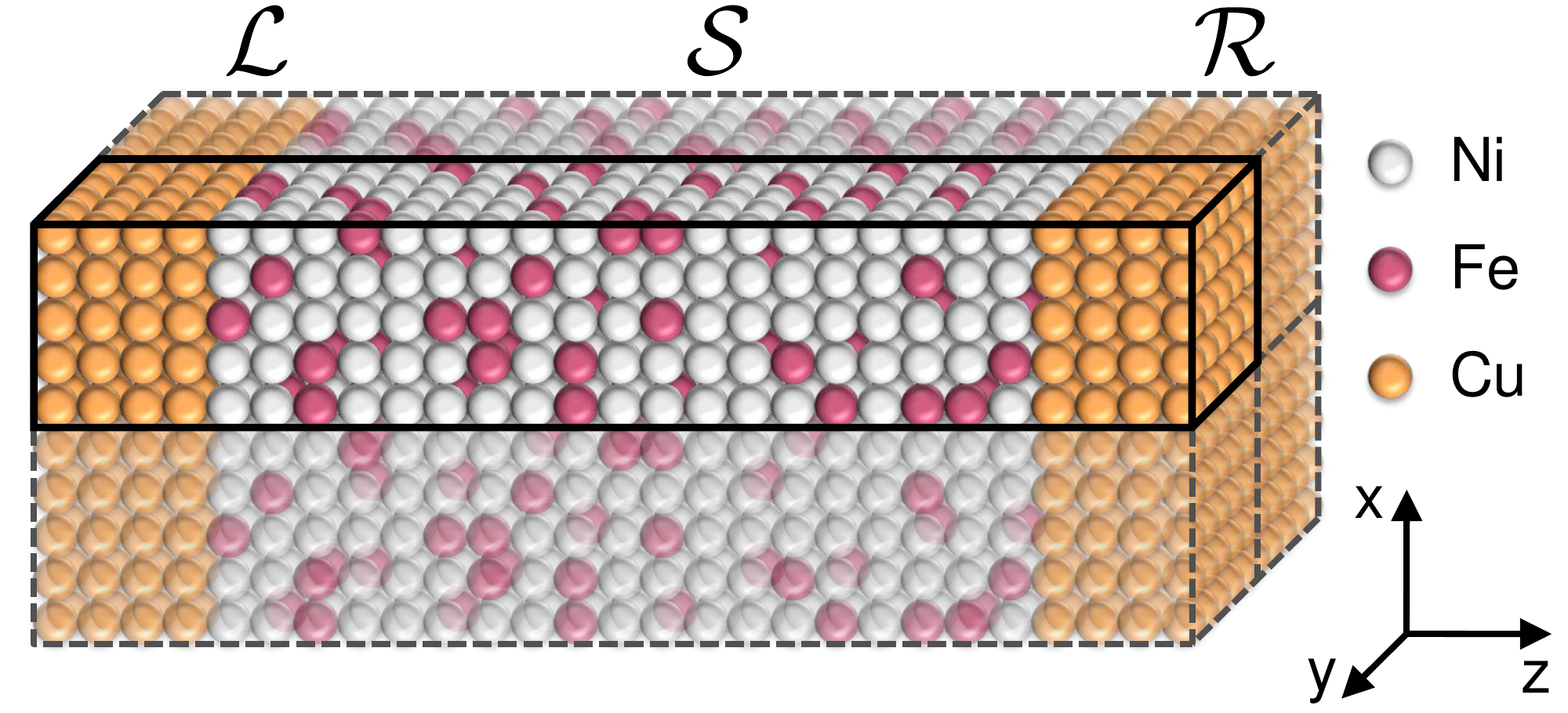} 
\end{center}
\caption{Example of a transverse supercell for a Cu$|$Py$|$Cu scattering geometry. Cu atomic layers form semi-infinite ballistic leads denoted $\mathcal{L}$ and $\mathcal{R}$. The scattering region $\mathcal{S}$ consists of a thickness $L$ of the substitutional disordered Ni$_{80}$Fe$_{20}$ alloy, Permalloy, sandwiched between the leads. Each atomic layer in $\mathcal{L}|\mathcal{S}|\mathcal{R}$ contains $5 \times 5$ atoms. The layers are parallel to the $xy$ plane and in the calculations this structure is repeated in the $x$ and $y$ directions so that an infinite periodic structure arises. 
}
\label{Fig2}
\end{figure}

The starting point for our determination of ${\bf j}_c$ and ${\bf j}_{s \alpha}$ is the solution of a single-particle Schr\"odinger equation 
\cite{footnote2} 
$H\Psi=E\Psi$ for a two terminal $\mathcal{L}|\mathcal{S}|\mathcal{R}$ configuration in which a disordered scattering region $\mathcal{S}$ is sandwiched between crystalline left- ($\mathcal{L}$) and right-hand ($\mathcal{R}$) leads, Fig.~\ref{Fig2}. 
The quantum mechanical calculations are based upon Ando's wave-function matching (WFM) \cite{Ando:prb91, Zwierzycki:pssb08} method formulated in terms of a localized orbital basis $|i\rangle$. Our implementation \cite{Xia:prb06, Starikov:prb18} is based upon a particularly efficient minimal basis of tight-binding muffin tin orbitals (TB-MTOs) \cite{Andersen:prl84, *Andersen:85, *Andersen:prb86} with $i=Rlm\sigma$ in combination with the atomic spheres approximation (ASA) \cite{Andersen:prb75}. Here $R$ is an atom site index and $lm\sigma$ have their conventional meaning. In terms of the basis $|i\rangle$, the wavefunction $\Psi$ is expressed as
\begin{equation}
\label{eq:psi}
|\Psi\rangle= \sum_i |i\rangle \langle i|\Psi\rangle
\end{equation}
and the Schr\"odinger equation becomes a matrix equation with matrix elements $\langle i |H| j \rangle$. $\Psi$ is a vector of coefficients with elements $\psi_i \equiv \langle i|\Psi\rangle $ extending over all sites $R$ and over the orbitals on those sites, for convenience collectively labelled as $i_R$. 

A number of approximations makes solution of the infinitely large system tractable. First, by making use of their translational periodicity, the WFM method eliminates the semiinfinite leads by introducing an energy dependent ``embedding potential'' on each atom in the layer of atoms bounding the scattering region \cite{Ando:prb91, Zwierzycki:pssb08}. 
Second, the system is assumed to be periodic in the directions transverse to the transport direction (taken to be the $z$-axis). This makes it possible to characterize the scattering states with a transverse Bloch wavevector ${\bf k}_{\parallel}$. Fixing ${\bf k}_{\parallel}$ and the energy (typically, but not necessarily, at $E=E_F$), the Schr\"odinger equation is first solved for each lead yielding several eigenmodes $\mu$ and their corresponding wavevectors $k_{\perp\mu}$. For propagating solutions $k_\perp$ must be real. By calculating the velocity vectors $\bf{v}$ for $\bf{k}_\mu \equiv (\bf{k}_\parallel,\bf{k}_{\perp\mu})$, propagating modes in both leads can be classified as right-going ``$\bf{v}^+$'' or left-going ``$\bf{v}^-$''. To simplify the notation, we rewrite $\mathbf{k}_\mu\equiv \mu\mathbf{k}$. The lead solutions are then used as boundary conditions to solve the Schr\"odinger equation in the scattering region for states transmitting from left to right $(L\rightarrow R)$ and right to left $(R\rightarrow L)$.
The complete wavefunction can be written as
\begin{equation}
\label{leftpsi}
\Psi^+_{\mu{\bf k}} =
\left( \begin{array}{c}
\Psi^{\mathcal{L}+}_{\mu{\bf k}} +\sum\limits_{\nu{\bf l}} r_{\nu{\bf l},\mu{\bf k}} \Psi^{\mathcal{L}-}_{\nu{\bf l}}  \vspace{0.5em}
\\
\Psi^{\mathcal{S}+}_{\mu{\bf k}}  \vspace{0.5em}
\\
\sum\limits_{\nu{\bf l}} t_{\nu{\bf l},\mu{\bf k}} \Psi^{\mathcal{R}+}_{\nu{\bf l}}
\end{array} \right)
\end{equation}
and
\begin{equation}
\label{rightpsi}
\Psi^-_{\mu{\bf k}} =
\left( \begin{array}{c}
\sum\limits_{\nu{\bf l}} t_{\nu{\bf l},\mu{\bf k}} \Psi^{\mathcal{L}-}_{\nu{\bf l}} \vspace{0.5em}
\\
\Psi^{\mathcal{S}-}_{\mu{\bf k}} \vspace{0.5em}

\\
\Psi^{\mathcal{R}-}_{\mu{\bf k}} + \sum\limits_{\nu{\bf l}} r_{\nu{\bf l},\mu{\bf k}} \Psi^{\mathcal{R}+}_{\nu{\bf l}}
\end{array} \right)
\end{equation}
where $t$ and $r$ are matrices of transmission and reflection probability amplitudes. Because the leads contain no disorder by construction, we will be focusing on the wave functions $ \Psi^{\mathcal{S}\pm}_{\mu{\bf k}}$ of \eqref{leftpsi} and \eqref{rightpsi} in the scattering region to calculate the current tensor  separately for $\Psi^{\mathcal{S}+}(L\rightarrow R)$ and $\Psi^{\mathcal{S}-}(R\rightarrow L)$ summed over all ${\mu{\bf k}}$. The former yields a right going electron current whereas the latter yields a left going hole current when an infinitesimal voltage bias is applied.

\subsection{Calculating the full current tensor}
\label{SSec:curr}


In this section we discuss a method to calculate from first principles charge and spin currents between atoms using localized orbitals. This is particularly suited for methods using the ASA and TB-MTOs \cite{Andersen:prl84, Andersen:85, Andersen:prb86}. In an independent electron picture \cite{footnote2}, the particle density is given by $n({\bf r},t)=|\Psi({\bf r},t)|^2$ where we omit the subscripts $\mu{\bf k}$ and superscripts $\pm$ of the previous section. Particle conservation requires that $\partial_t n({\bf r},t) + \nabla \cdot {\bf j}({\bf r})=0$ where ${\bf j}({\bf r})$ is the probability current density.  A volume integral over the atomic sphere (AS) $S_P$ centered on atom $P$ yields 
\begin{equation}\label{eq:continuity}
\partial_t n_P=-\iint \limits_{S_P}\fat{j}\cdot d\fat{S}
\end{equation}
where $n_P$ is the number of particles in the AS that can only change in time if a current flows in or out of the atomic sphere. The ASA requires filling all of space with atomic Wigner Seitz spheres and leads to a discretized picture in which the net current into or out of $S_P$ is balanced by the sum of currents leaving or entering the neighbouring atomic spheres. This interpretation works especially well when we use TB-MTOs whose hopping range is limited to second or third nearest neighbors \cite{Andersen:prl84, Andersen:85, Andersen:prb86}. 

The coefficients in $\Psi$ relating to the basis on atom $P$ can be labelled $\Psi_P$
\begin{equation}
   | \Psi_P \rangle = {\widehat P} |\Psi \rangle \,,
\end{equation}
with
\begin{equation}
{\widehat P} =  \sum_{i_P} \left|i_P \rangle \langle i_P \right| \,.
\end{equation}
The number of electrons $n_P$ on atom $P$ is then defined as
\begin{equation}
n_P = \langle \Psi |{\widehat P}| \Psi \rangle \equiv \langle \Psi_P | \Psi_P \rangle
\end{equation}
where the bra-ket notation implies an inner product.
We denote the net current from atom $Q$ to atom $P$ with $j_c^{PQ}$, measured in units of the electron charge $-e$ where $e$ is a positive quantity. A sub-block of the Hamiltonian containing the hopping elements from atom $Q$ to atom $P$ is denoted $H_{PQ}$.

Similarly, the $\alpha$ component of the spin density on atom $P$ is
\begin{equation}
	s_{\alpha,P} = \langle \Psi_P | \sigma_{\alpha} | \Psi_P \rangle
\end{equation}
where $\sigma_{\alpha}$ is a Pauli matrix. For convenience we divide the spin density by $\hbar/2$ and express it as a particle density.
We also express the spin current as a particle current.
$j_{s\alpha}^{PQ}$ is the spin transfer into atomic sphere $P$ carried by electrons hopping from atom $Q$. It must be clear that it is the spin current exactly at the sphere boundary of atom $P$ and that the index $Q$ merely indicates from where the electrons hopped.

\subsubsection{Interatomic electron currents}
\label{electron}
With the above definitions, we can rewrite the charge conservation equation \eqref{eq:continuity} as
\begin{equation} 
\label{dtnr}
\partial_t n_P = \sum_Q j_c^{PQ} \left(\Psi_P, \Psi_Q \right)
\end{equation}
where $j_c^{PQ}\left( \Psi_P, \Psi_Q \right)$ should change sign if $P$ and $Q$ are interchanged; the current from $Q$ to $P$ is minus the current from $P$ to $Q$. The current $j_c^{PQ}$ cannot depend on electron densities located elsewhere than on $Q$ or $P$ in an independent electron picture. Note that $j_c^{PP}=0$ in accordance with particle conservation. In the Schr\"odinger picture we have
\begin{equation}
\partial_t \Psi = \frac{1}{i\hbar} H \Psi \,.
\end{equation}
From this we can deduce that with any general time-dependent wavefunction $\Psi$ at a specific moment in time, the number of electrons on atom $P$ changes with the following rate
\begin{subequations}
\begin{align}
\partial_t n_P 
&=\big\langle \Psi \big| {\hat P} \big| \partial_t \Psi \big\rangle + 
  \big\langle \partial_t\Psi \big| {\hat P} \big | \Psi \big\rangle  \\
=\frac{1}{i\hbar} &\sum_Q { \bigg[ \big\langle \Psi_P \big| H_{PQ} \big| \Psi_Q \big\rangle - \big\langle \Psi_Q \big| H_{QP} \big| \Psi_P \big\rangle \bigg] }\,
\end{align}
\end{subequations}
which has the form of \eqref{dtnr} with 
\begin{equation} \label{je}
j_c^{PQ} = \frac{1}{i\hbar}  \bigg[ \big\langle \Psi_P  \big| H_{PQ}  \big| \Psi_Q  \big\rangle -  \big\langle \Psi_Q  \big| H_{QP}  \big| \Psi_P  \big\rangle \bigg] .
\end{equation}
It is easy to see from this expression that solving the time-independent Schr\"odinger equation $ H \Psi = E \Psi $ makes sure the charge on an atom stays constant. This formula can be used to calculate interatomic electron currents.

\subsubsection{Interatomic spin currents}
\label{spin}

The general form of the time dependence of the spin density on atom $P$ is similar to \eqref{dtnr}
\begin{equation}
\label{dtsr}
\partial_t s_{\alpha,P} = \sum_Q j_{s\alpha}^{PQ}\left(\Psi_P, \Psi_Q \right) \,,
\end{equation}
because spin is carried by electrons. $j_{s\alpha}^{PQ}$ is now not required to change sign if $Q$ and $P$ are interchanged because spin is not conserved \cite{footnote3}; it changes due to exchange torque as well as spin-orbit torque. This also means that $j_{s\alpha}^{PP}$ need not be zero and in fact it is the local torque on the spin density at $P$. Physically the rate of change of the total spin in a certain region consists of two contributions: the net spin flow into the region and a local torque, i.e.
\begin{equation}
\label{spconserv}
\partial_t s_{\alpha,P}=-\iint \limits_{S_P}\fat{j}_{s\alpha}\cdot d\fat{S}+\tau_{\alpha,P}.
\end{equation}
The general form of \eqref{dtsr} is consistent with the spin conservation equation \eqref{spconserv}.

From the Schr\"odinger equation we calculate the rate of change of spin on atom $P$ to be
\begin{subequations}
\begin{align}
\partial_t s_{\alpha,P} &=
\big\langle \Psi_P \big| \sigma_\alpha \big| \partial_t \Psi_P \big\rangle +
\big\langle \partial_t \Psi_P \big| \sigma_\alpha \big| \Psi_P \big\rangle  \\
= \frac{1}{i\hbar} & \sum_Q \bigg[ \big\langle \Psi_P \big| \sigma_\alpha H_{PQ} \big| \Psi_Q \big\rangle - \big\langle \Psi_Q \big| H_{QP} \sigma_\alpha \big| \Psi_P \big\rangle \bigg] \,
\end{align}
\end{subequations}
which has the form of \eqref{dtsr} with
\begin{equation} \label{js}
j_{s\alpha}^{PQ} = \frac{1}{i\hbar}  \bigg[ \big\langle \Psi_P \big| \sigma_\alpha H_{PQ} \big| \Psi_Q \big\rangle \! - \big\langle \Psi_Q \big| H_{QP} \sigma_\alpha \big| \Psi_P \big\rangle \bigg]. 
\end{equation}
If basis functions are defined within the ASA it is very clear that this is the spin current exactly at the sphere boundary of atom $P$ if $Q\neq P$ and it is the local torque if $Q=P$. 

As mentioned above, the change of spin in a sphere is the local torque $j_{s\alpha}^{PP}$ plus the sum of all spin currents $j_{s\alpha}^{PQ}$ into the sphere. The spin current leaving sphere $Q$ is not the same as the spin current entering sphere $P$, i.e. $j_{s\alpha}^{PQ} \neq -j_{s\alpha}^{QP}$ because spin is not conserved. This means there must also be torques acting on spins when they are ``between'' the atoms in addition to the torques inside the spheres. A torque is of course equal to the rate of change of spin. It can be relevant to compare this way of calculating the local torques to other methods \cite{Shi:prl06}.

\subsection{Layer averaged current tensor}
\label{SSec:prac}

\begin{figure}
\includegraphics[width=5 cm]{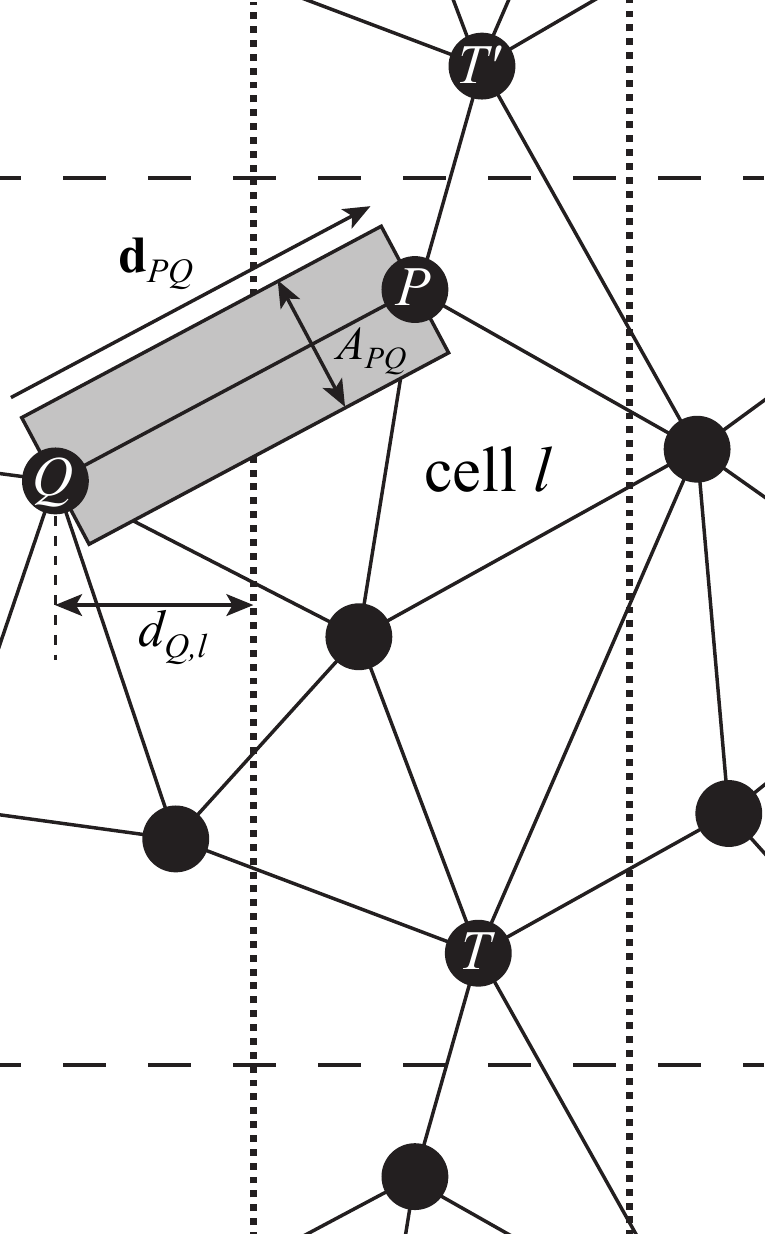} 
\caption{Illustration of a number of concepts defined in the text. Current flow is in the $z$ direction from left to right. The horizontal dashed lines indicate the (lateral) unit cell boundaries. The vertical dotted lines indicate layer boundaries in the $z$ direction. The gray area is a ``wire'' with assumed homogeneous current density that substitutes for the general spatial distribution of the current between $Q$ and $P$, which can therefore be left unknown.}
\label{Fig3}
\end{figure}

The information obtained from the calculations outlined in the previous section has the form of a network flow or a weighted graph. Every node in the graph represents an atom and each end of a connection is accompanied by 4 numbers representing currents. These currents can be arranged in a 4-vector for convenience: $ {\bf j}^{PQ} = (j_c^{PQ}, j_{sx}^{PQ}, j_{sy}^{PQ}, j_{sz}^{PQ} ) $. The problem we now address is how to convert this information to a continuum current density tensor represented on a discrete grid. We start by separating the system into layers $l$. If there is periodicity in the $x$ and $y$ directions (or if the system is finite) this will define cells with volumes $V_l$ depending on the thicknesses of the layers. If  there is periodicity in the $xy$ plane, we need to characterize equivalent atoms $T$ and $T'$ by the unit cell ${\bf R}$ they are in, in order to know in which direction an interatomic current is flowing, see \cref{Fig3}. That can be done by decomposing the Hamiltonian
\begin{equation}
H= \sum_{\fat{R}} H_{\fat{R}} e^{i\fat{k\cdot R}}
\end{equation}
and calculating the currents, e.g. ${\bf j}^{PT}$ and ${\bf j}^{PT'}$, for each term separately. We label every atom in the unit cell and every relevant translation of it with a different index ($P$ or $Q$ here). Note that in ${\bf j}^{PQ}$ every atom $P$ lies inside the original unit cell; $Q$ can be either inside or outside. This way we are sure that we count all the currents that should be attributed to one unit cell exactly once. Details of how a current ${\bf j}^{PQ}$ is distributed in space are not known so we imagine that the flow is homogeneous in a wire with arbitrary cross-section $A_{PQ}$ and volume $V_{PQ} = A_{PQ}|{\bf d}_{PQ}|$, where ${\bf d}_{PQ}$ is the vector pointing from atom $Q$ to $P$.

The current density tensor integrated over the volume of this wire is ${\stackrel{\leftrightarrow}{j}}\!^{PQ} V_{PQ} = {\bf j}^{PQ} \otimes {\bf d}_{PQ} $ and does not depend on the cross-section. The average current density tensor times the volume of cell $l$, ${\stackrel{\leftrightarrow}{j}}\!_l V_l$, is now the sum of current densities of all these wires integrated within cell $l$. We define a parameter that indicates how much of the wire $PQ$ lies outside the cell at the atom $Q$ end
\begin{equation}
\beta_{QP,l} = \begin{cases}
0 & \mbox{if  $Q$ inside cell $l$} \\
d_{Q,l}/d_{z,QP} &\mbox{if  $Q$ outside cell $l$}
\end{cases}
\end{equation}
where $d_{P,l}$ is the $z$-distance from atom $P$ to the closest boundary plane of layer $l$.
Since the spin current changes between $Q$ and $P$, we make a linear interpolation
\begin{equation}
{\bf j}^{PQ}(c) = c \,{\bf j}^{PQ} - (1-c) {\bf j}^{QP} \,,
\end{equation}
where $c$ is a parameter that runs from 0 to 1 depending on the position between $Q$ and $P$. Now the part of ${\stackrel{\leftrightarrow}{j}}\!^{PQ} V_{PQ}$ 
that should be counted into ${\stackrel{\leftrightarrow}{j}}\!_l V_l$ is
\begin{eqnarray}
\int^{1-\beta_{PQ,l}}_{\beta_{QP,l}} &&{\bf j}^{PQ}(c)\otimes {\bf d}_{PQ} \, {\rm d}c  = \nonumber \\
\tfrac{1}{2} &&\left[ \left( 1-\beta_{PQ,l}\right)^2 - \left(\beta_{QP,l}\right)^2\ \right] {\bf j}^{PQ}\otimes {\bf d}_{PQ} \nonumber \\
+\tfrac{1}{2} &&\left[ \left(1-\beta_{QP,l}\right)^2 - \left(\beta_{PQ,l}\right)^2\ \right] {\bf j}^{QP}\otimes {\bf d}_{QP} \,.
\end{eqnarray}
Note that ${\bf d}_{QP} = - {\bf d}_{PQ}$. The average current density tensor in cell $l$ is then
\begin{equation}
\label{jten}
\stackrel{\leftrightarrow}{j_l} =
\frac{1}{V_l} \sum_{P,Q} \tfrac{1}{2} 
\left[ \left(1-\beta_{PQ,l}\right)^2 - \left(\beta_{QP,l}\right)^2\ \right] 
                {\bf j}^{PQ}\otimes {\bf d}_{PQ} .
\end{equation}
Now we can multiply with the cross-sectional area of the unit cell to obtain a total current per unit voltage between two leads that can be compared directly with the total Landauer-B\"uttiker conductance. This is an important criterion to verify the numerical implementation of the above local current scheme. Eventually, the current density tensor is divided by the total conductance or total current to yield normalised current densities that will be presented in Sec.~\ref{Sec:calc}.

\subsection{First principles calculations}
\label{SSec:scatcalc}

The formalism for calculating currents sketched in the previous section has been applied to the wave functions \eqref{leftpsi} and \eqref{rightpsi} expanded in a basis of TB-MTOs. We here briefly recapitulate some technical aspects of the TB-MTO-WFM method \cite{Xia:prb06, Starikov:prb18} that need to be checked in the scattering calculations to determine the dependence of the spin currents and quantities derived from the spin currents. 

TB-MTOs are a so-called ``first-principles'' basis constructed around partial waves, numerical solutions at energy $E$ of the radial Schr\"odinger equation for potentials that are spherically symmetric inside atomic Wigner-Seitz spheres (AS). The MTOs and matrix elements of the Hamiltonian are constructed from AS potentials calculated self-consistently within the DFT framework combined with short-range ``screened structure constants'' \cite{Andersen:prl84, *Andersen:85, *Andersen:prb86}. Inside an AS, the MTO is expressed as products of partial waves, spherical harmonics and spinors so that a MTO is labelled $| R lm \sigma \rangle$ in the notation of Sec.~\ref{SSec:scatt}.

%


\subsubsection*{SOC: two and three center terms}

Spin-orbit coupling is included in a perturbative way by adding a Pauli term to the Hamiltonian \cite{Andersen:prb75, Brooks:prl83, Daalderop:prb90a, Starikov:prb18}.
TB-MTOs lead to a Hamiltonian with one, two and three centre tight-binding-like terms where the three-centre SOC terms introduce longer range hopping \cite{Starikov:prb18} than the next-nearest neighbour interaction of the ``screened structure constant matrix'' \cite{Andersen:prl84, *Andersen:85, *Andersen:prb86}. Explicit calculation demonstrated that omitting these terms had negligible effect on the resistivity and Gilbert damping but reduced the computational cost by some 70\% \cite{Starikov:prb18}. Unless stated otherwise, calculations will only include two center terms. 


\subsubsection*{Partial wave expansion}

In the TB-MTO-WFM code \cite{Xia:prb06, Starikov:prb18} the wavefunctions inside atomic spheres are expanded in a partial wave basis that is in principle infinite. In practice the infinite summation must be of course be truncated. For transition metal atoms, we usually use a basis of $spd$ orbitals and test the convergence with an $spdf$ basis. Unless stated otherwise, an $spd$ basis will be used.  


\subsubsection*{Scattering configuration: lateral supercells}

Transport in ballistic metals can be studied by constructing an $\mathcal{L}|\mathcal{S}|\mathcal{R}$ scattering configuration with $1 \times 1$ periodicity perpendicular to the transport direction and exploiting the periodicity of the system. Because systems with thermal and chemical disorder or multilayers are not periodic, we model them with a scattering region consisting of a large unit cell transverse to the transport direction that we call a ``lateral supercell'', Fig.~\ref{Fig2}. Typically this consists of $N \times N$ primitive $1 \times 1$ unit cells containing $M=N^2$ atoms. No periodicity is assumed in the transport direction itself that is typically $L$ atomic layers in length \cite{Xia:prb06, Starikov:prb18}. The size of supercell that can be handled is constrained by computational expense. This scales as the third power of the number of atoms in a lateral supercell and linearly in the length of the scattering region, as $M^3 L = N^6L$. The lateral supercell leads to a reduced two-dimensional (2D) Brillouin zone (BZ) and a saving on the BZ sampling so that the computational effort ultimately scales as $M^2 L = N^4 L$. An alloy like Py has no long-range order, thus the supercell approximation is only exact for infinite supercell size. In practice, it will turn out that very good results can be obtained for both Pt and Py using remarkably small lateral supercells.

The simplest way to perform scattering calculations for e.g. thermally disordered Pt is to use ballistic Pt leads. We will examine the effect of a different choice of lead material on the parameter estimates by using other lead materials.
The lattice constants of Au ($a=4.078 \AA$) and Ag ($a=4.085\AA$) are much closer to that of Pt ($a=3.923\AA$) than is that of Cu ($a=3.615\AA$) and by compressing them slightly, they can be made to match Pt without significantly changing their electronic structures. The requirement that leads should have full translational symmetry precludes using an alloy as a lead material \cite{Starikov:prb18}. To study the properties of Py ($a=3.541\AA$), it is convenient to use slightly compressed Cu as lead material. To use Cu as a lead for Pt (as mentioned in Sec. \ref{Sec:Intro}), we constructed a relaxed Cu$|$Pt$|$Cu scattering configuration by choosing appropriately matched supercells for Cu and Pt. As long as we are only interested in the bulk properties of Pt and Py, the choice of lead material should not matter; we will demonstrate this explicitly. 


\subsubsection*{Alloy disorder}

Disordered substitutional alloys can be modelled in lateral supercells by randomly populating supercell sites with AS potentials subject to the constraint imposed by the stoichiometry of the targeted experimental system. In principle, the AS potentials can  result from self-consistent supercell calculations. In practice, we use the very efficient coherent-potential-approximation (CPA) \cite{Soven:pr67} implemented with TB-MTOs \cite{Turek:97} to calculate optimal Ni and Fe potentials for Permalloy. Since we will not be studying interface properties in this paper, we will use CPA potentials calculated for bulk Py rather than using a version of the CPA generalized to allow the optimized potentials to depend on the layer position with respect to an interface \cite{Turek:97}.  

\subsubsection*{Thermal disorder}

 Many experiments in the field of spintronics are performed at room temperature where  transport properties are dominated by temperature induced lattice and spin disorder. We will model this type of disorder within the adiabatic approximation using a recently developed ``frozen thermal disorder scheme'' \cite{LiuY:prb11, LiuY:prb15}. In Ref.~\onlinecite{LiuY:prb15} correlated atomic displacements were determined from the results of lattice dynamics calculations by taking a superposition of phonon modes weighted with a temperature dependent Bose-Einstein occupancy; this was shown to very satisfactorily reproduce earlier results obtained in the lowest order variational approximation (LOVA) with electron phonon matrix elements calculated from first principles with linearized MTOs \cite{Savrasov:prb96b}. Rather than trying to extend this ab-initio approach to disordered alloys, we adopt the simpler procedure of modelling atomic displacements with a Gaussian distribution \cite{LiuY:prb11} and choosing the root-mean square displacement $\boldmath{\Delta}$ to reproduce the experimental resistivity \cite{LiuY:prb15}. Here, it is important to note that $\Delta$ can depend on the choice of orbital basis, supercell size and inclusion of three center terms. For RT Pt, $\Delta$ is chosen to yield the room temperature resistivity  $\rho_{\rm Pt} = 10.8~\mu\Omega$~cm \cite{HCP90}. With this approach, the results we obtain for $l_{\rm sf}$ and $\Theta_{\rm sH}$ for RT Pt differ slightly from our earlier work \cite{LiuY:prb15, WangL:prl16}. However, because Pt satisfies the Elliot-Yafet relationship, the products $\rho \, l_{\rm sf}$ and $\sigma \, \Theta_{\rm sH}$ agree with those earlier publications. Here $\sigma = 1/\rho$ is the conductivity.
 
Spin disorder is treated analogously \cite{LiuY:prb11}. Because spin-wave theory underestimates the temperature induced magnetization reduction, we choose a Gaussian distribution of polar rotations and a uniform distribution in the azimuthal angle to reproduce the temperature dependent magnetization \cite{LiuY:prb15, Starikov:prb18}. The lattice disorder is then chosen so that spin and lattice disorder combined reproduce the experimental \cite{Ho:jpcrd83} resistivity of Py, $\rho_{\rm Py} = 15.4~\mu\Omega$~cm, at 300 K. 

For both lattice and spin disorder it is necessary to average over a sufficient number of configurations of disorder and to study the effect of the supercell size. All results in this paper are averaged over 20 configurations of disorder.

\subsubsection*{k-point sampling}

To count all possible scattering states at the Fermi energy a summation over the Bloch wavevectors ${\bf k}_\parallel$ in the 2D BZ common to the real space supercells must be performed. We sample the BZ uniformly dividing each reciprocal lattice vector into $Q$ intervals. For an $N \times N$ real space lateral supercell, sampling the 2D BZ with $Q \times Q$ k-points leads to a sampling that is equivalent to an $NQ \times NQ$ sampling for the primitive $1 \times 1$ unit cell.

\subsubsection*{Slab length}

To extract a value of the SDL characteristic of the bulk, it is important to verify that the decay of the spin current is exponential over a length at least several times longer than $l_{\rm sf}$ and independent of the lead materials. Because the  bulk material is always embedded between two ballistic leads, a deviation from  exponential behavior is unavoidable close to the interfaces. We will see that acceptable exponential behaviour is obtained if the lateral supercell and k-space sampling are sufficiently large and the scattering region is sufficiently long. 

\subsubsection*{Averaging L$\rightarrow$R and R$\rightarrow$L currents}

At an interface, the wave character of particles in a quantum mechanical calculation leads to interference between the incident and reflected waves and we observe standing waves in the spin currents that decay away from the interface. These fluctuations are largest close to the left interface for ${\bf j}_{s\alpha}^{LR}(z)$ and to the right interface for ${\bf j}_{s\alpha}^{RL}(z)$ and gradually disappear towards the other interface, largely paralleling the corresponding unscreened particle accumulations $n^{LR}(z)$ and $n^{RL}(z)$. Even though the oscillations are real effects, we are interested in comparing our data with semiclassical descriptions that do not contain them. 
For an ideal bulk system, the spin current ${\bf j}_{s\alpha}^{LR}(z)$ accompanying a current of electrons from left to right should be identical to the spin current ${\bf j}_{s\alpha}^{RL}(z)$ arising from passing a current of holes from right to left. 
In order to extract various bulk parameters, we use the unscreened particle accumulations $n^{LR}(z)$ and $n^{RL}(z)$ in the following expression to reduce the fluctuations
\begin{equation}
\label{scrave}
{\bf j}_{s\alpha}^{\rm av} = 
       \frac{n^{RL}}{n^{LR} + n^{RL}} {\bf j}_{s\alpha}^{LR}  
     + \frac{n^{LR}}{n^{LR} + n^{RL}} {\bf j}_{s\alpha}^{RL} \, .
\end{equation}
All results presented in this publication are based upon such averaging. 

\subsubsection*{Spin polarized leads}

\begin{figure}[b]
	\includegraphics[width=8.4 cm]{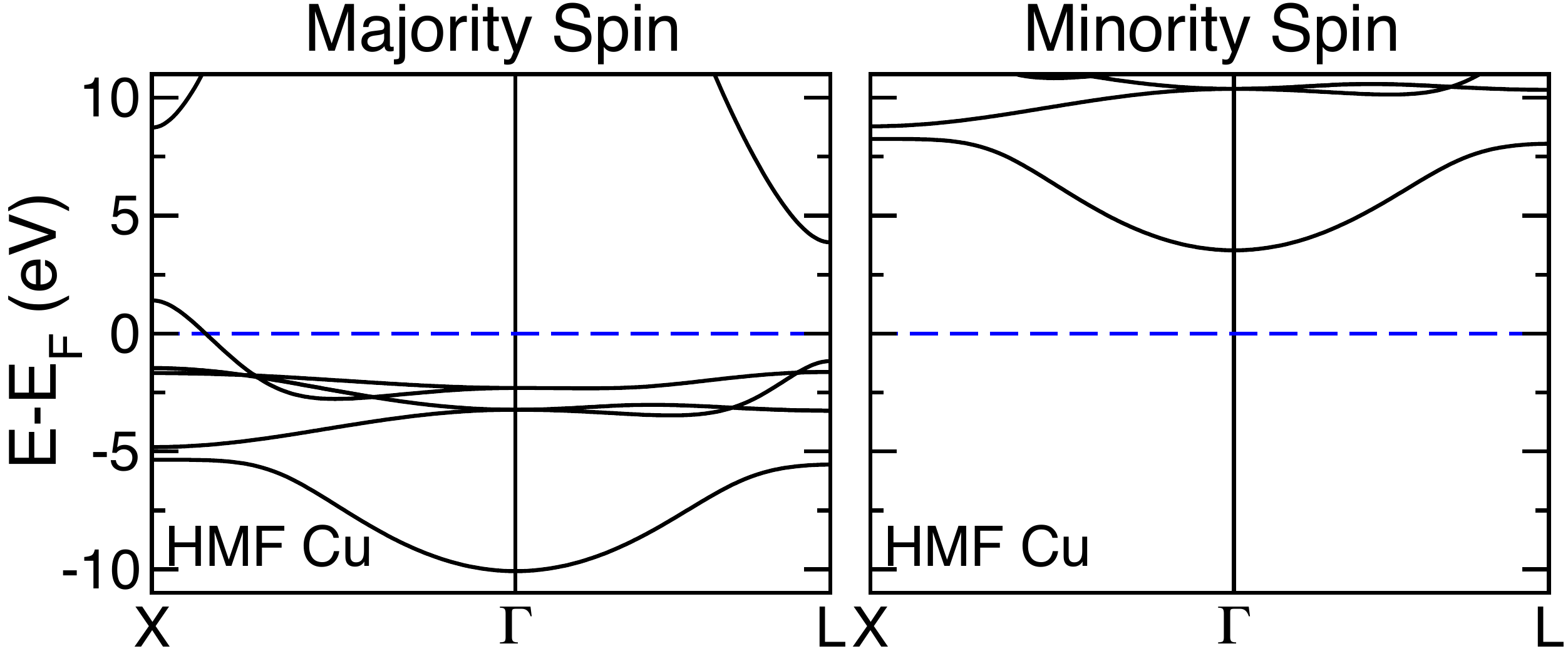} 
	\caption{Majority (lhs) and minority (rhs) spin band structures of Cu when a repulsive constant potential of 1 Rydberg is added to the minority spin potential. The effect is to remove all minority spin states from the Fermi energy.}
	\label{Fig4}
\end{figure}

In order to study the SDL of a material, e.g. Pt, we need to attach magnetic leads to it to inject a spin polarized current. The polarization of a magnetic lead will in general not be unity and the lead$|$Pt interface will result in a loss of spin signal entering Pt. We maximise the incident spin current by making a halfmetallic ferromagnet (HMF) out of a noble metal. To do so, we add a constant to the potential of one spin channel of the lead material in the scattering calculation to remove that spin channel from the Fermi energy entirely. This is illustrated in Fig.~\ref{Fig4} where a constant of one Rydberg has been added to the ``minority'' spin potential to make Cu HMF. Since we are not interested in interface properties in the present publication, it is of no concern that this potential is not self-consistent. In a study of real interfaces, more attention would need to be paid to this issue \cite{Gupta:tbp19}. We denote Cu made to be HMF in this way as Cu$\uparrow$.

\section{Results}
\label{Sec:calc}

We illustrate the spin-current formalism with calculations of the SDL $l_{\rm sf}$ for Pt and Py, the current polarization $\beta$ for Py and the spin Hall angle $\Theta_{\rm sH}$ for Pt, all at room temperature. The words spin currents and spin current densities will be used interchangeably. Because the results of calculations are always presented in terms of spin current densities normalized with respect to the constant total current $j\equiv j_c^z(z)$ in the $z$ direction, we omit the $\widehat{\phantom{j}}$ over $j_s(z)$ in \eqref{eq:jss} when there is no ambiguity. 

\subsection{${\bf l_{\rm sf}}$ for Pt}
\label{SSec:lsf_Pt}

We inject a fully polarized current from a HMF ballistic Au$\uparrow$ lead into RT thermally disordered Pt along the $z$-axis chosen to be the fcc (111) direction perpendicular to close packed atomic layers with the spin current polarized along the $z$-axis. The distribution of the random displacments of the Pt atoms from their equilibrium lattice positions is Gaussian  with a rms displacement $\Delta$ chosen to reproduce the experimental RT resistivity.

\begin{figure}[b]
\includegraphics[width=8.5 cm]{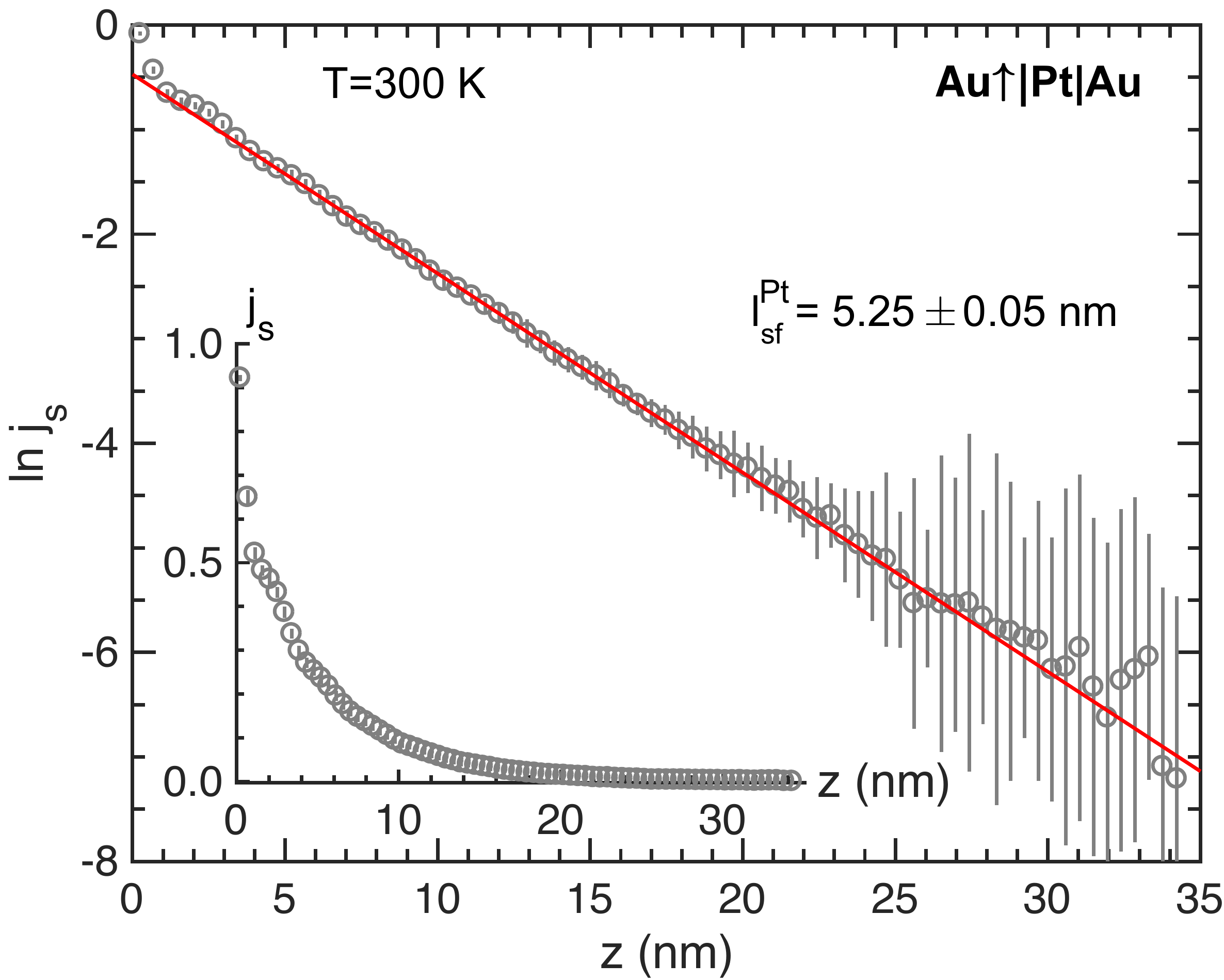} 
\caption{Natural logarithm of the spin current injected into RT Pt as a function of the coordinate $z$ in the transport direction. The inset shows the spin current on a linear scale. The current was extracted from the results of a scattering calculation for a two-terminal Au$\uparrow$$|$Pt$|$Au configuration using a $7 \times 7$ lateral supercell. The red line is a weighted linear least squares fit; the error bar in the value $5.25\pm0.05$ results from different ``reasonable'' weightings and cutoff values. 
}
\label{Fig5}
\end{figure}

The natural logarithm of $j_s(z)$ is shown in Fig.~\ref{Fig5}. In the linear plot shown in the inset, we see an initial rapid decrease of the spin current over a distance of order 1~nm from a value close to unity at the interface, followed by oscillatory damped behaviour that rapidly decays to 0. The exponential decay over almost five orders is very clear in the logarithmic plot. The red line is a weighted linear least squares fit to \eqref{eq:js} from which data up to 4 nm are excluded (including the interface and first half cycle of the  oscillatory term). The slope directly yields a value of $l_{\rm sf}^{\rm Pt} = 5.25 \pm 0.05 \,$nm. The weights are selected to be the inverse of the variance of the spin currents that results from 20 different configurations of thermal disorder. 
The error bar is then estimated using weighted residuals.

The initial decrease at the interface of $\sim e^{-\frac{1}{2}}$ over a length of $z = 1 \,$nm leads directly to an ``interface'' $l_{\rm sf}^I \sim 2 \,$nm. Using the definition \cite{Baxter:jap99, *Park:prb00, *Eid:prb02} of the interface ``spin memory loss'' parameter $\delta= t_I/{l_{\rm sf}^I}$ in terms of an interface thickness $t_I=1\,$nm yields a value of $\delta \sim 0.5$, a reasonable value \cite{Bass:jpcm07}. The clearly visible oscillations in the spin current are not predicted by semiclassical treatments. We attribute them to Fermi surface nesting-like features but more analysis would be required to establish this firmly.

The results shown in Fig.~\ref{Fig5} were calculated in a $7 \times 7$ Pt lateral supercell with an $spd$ basis and using a 2D BZ sampling of $32 \times 32$ k points equivalent to a $224 \times 224$ sampling for a $1 \times 1$ unit cell. In the remainder of this section we will examine how $l_{\rm sf}^{\rm Pt}$ depends on these and a number of the other computational parameters discussed in Sect.~\ref{SSec:scatcalc}. 

\subsubsection*{Supercell size}
\label{subsecA:supercell}

\begin{figure}[b]
\centering
\includegraphics[width=8.5 cm]{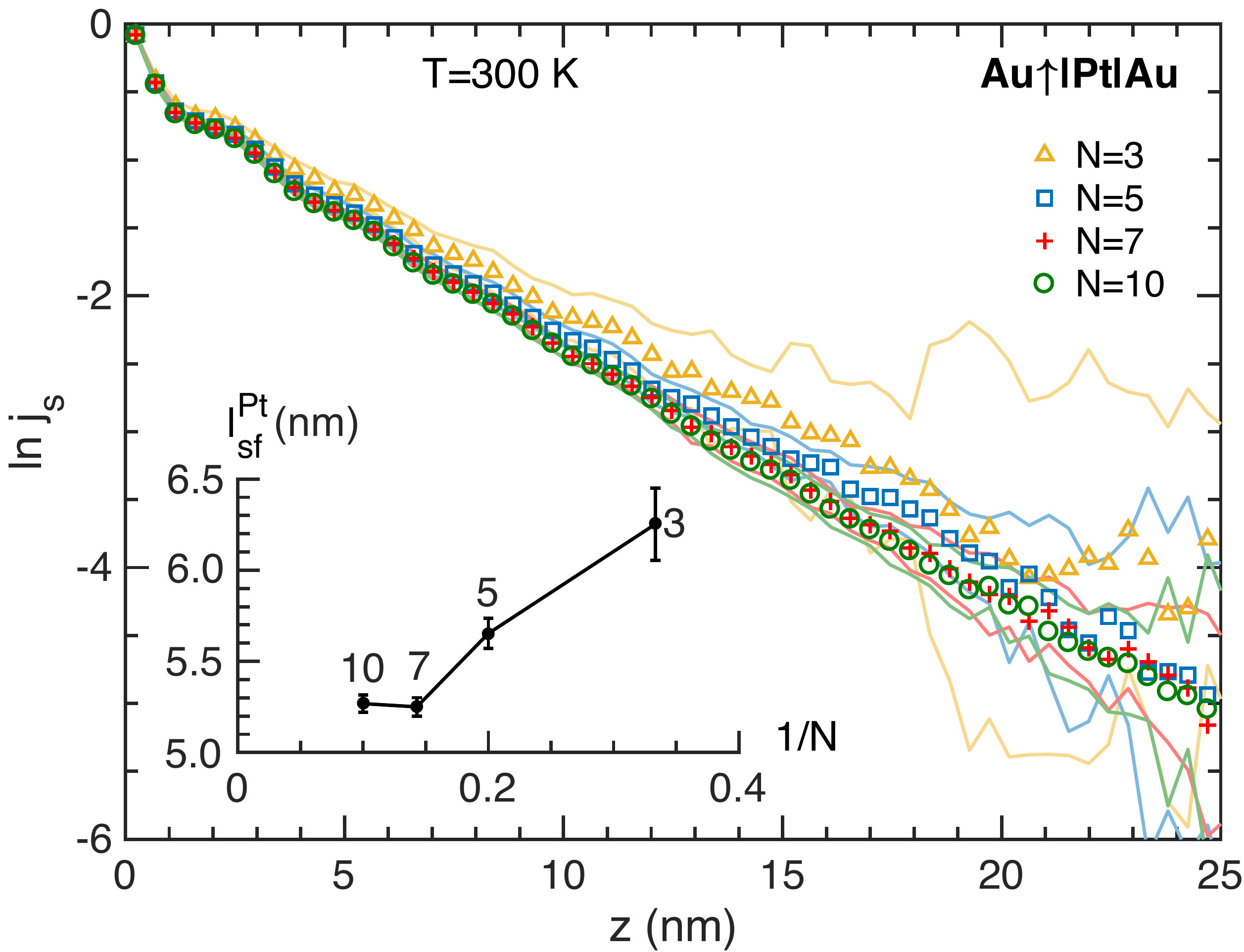}
\caption{Natural logarithm of the spin current density vs length for various $N \times N$ supercells ($N=3,5,7,10$) for a 35 nm long Pt slab at 300 K. The color coordinated symbols and solid lines indicate the mean and a measure of the spread of the data from 20 different configurations, respectively. Inset: SDL obtained from the linear fit of $\ln j_s(z)$ shown as a function of $1/N$. The numerical values of $l_{\rm sf}$ are given in Table~\ref{tab:SCPt}. 
}
\label{Fig6}
\end{figure}

It is not a priori clear how large a lateral supercell should be in order to adequately represent diffusive transport. On the one hand, one might expect it should be larger than the mean free path; in that case, this project would be doomed to failure for all but the most resistive of materials. On the other hand, only electrons scattered through $90^\circ$ ``know'' about the lateral translational symmetry. In Fig.~\ref{Fig6}, we show the natural logarithm of the normalized spin current density calculated for a Au$\uparrow$$|$Pt$|$Au scattering geometry using Pt $N \times N$ lateral supercells with $N=3, 5, 7, 10$; the largest supercell contains some 15000 atoms. For each value of $N$, we choose the BZ k-sampling parameter $Q$ so that  $NQ \sim 160$ in order to maintain a constant reciprocal space sampling equivalent to $160 \times 160$  for a 1$\times$1 primitive unit cell. The main features seen in Fig.~\ref{Fig5} are reproduced for all values of $N$. The most important trend is that $l_{\rm sf}$ decreases slightly with increasing $N$. As seen clearly in the inset, it converges rapidly to a value of $\sim 5.25\,$nm; the values are given separately in \cref{tab:SCPt}. For room temperature Pt, we see that it is sufficient to use a 7$\times$7 supercell.

Perhaps more striking is how rapidly the error bar decreases; see the inset. This can be easily understood. In an $N \times N$ lateral supercell, a single configuration of disorder ``seen'' by a spin before it flips contains of order $N^2 l_{\rm sf}/d$ atoms where $d$ is the spacing between Pt (111) planes $\sim 0.2\,$nm. For $N=3$, this amounts to only about 250 atoms, for $N=10$, it is about 2500. For short values of $l_{\rm sf}$ or small lateral supercells, we expect very large configuration to configuration variation and to have to include more configurations of disorder in our configuration averaging. By itself, this will not be sufficient because the freedom available to sample thermal disorder in a small supercell is intrinsically limited e.g. long wavelength transverse fluctuations cannot be represented in small supercells.


\begin{table}[t]
\caption{Dependence of the calculated SDL of RT Pt on the $N \times N$ supercell size for $N=3,5,7,10$. Calculations were performed with a k-point sampling equivalent to $160\times160$ for a $1\times1$ supercell in each case.}
\begin{ruledtabular}
\begin{tabular}{rccc}
$N$	& \multicolumn{1}{c}{$spd$ + 2 center} 
                       & \multicolumn{1}{c}{$spd$ + 3 center}
                                        & \multicolumn{1}{c}{$spdf$ + 2 center}\\
			\hline
3   & $6.25 \pm 0.20$  &  				& 				\\
5   & $5.65 \pm 0.08$  & $5.22\pm0.09$ 	& $5.21\pm0.07$	\\    
7   & $5.25 \pm 0.05$  & $4.96\pm0.07$	&				\\    
10  & $5.27 \pm 0.05$  & 				&				\\    
\end{tabular}
\end{ruledtabular}
\label{tab:SCPt}  
\end{table}

This has another important consequence. If we assume that a Au$\uparrow$ lead has a single scattering state per ${\bf k}_\parallel$ point in a $1 \times 1$ primitive interface unit cell, this means we begin with $160 \times 160 =25600$ states incident on the scattering region. 
For $z=6 \, l_{\rm sf} = 31 \,$nm, $e^{-6} \sim= \frac{1}{400}$. Of the $25600$ scattering states we started with in the left hand lead, we lose half at the interface and eventually only about 32 states are transmitted into the right hand lead without flipping their spins. This accounts for the large amount of noise seen in the spin current density for large values of $z$. This can be reduced to some extent by increasing the number of k points used to sample the BZ but is ultimately limited by a too-small supercell size.

\subsubsection*{k-point sampling}
\label{subsec:k sampling}

\begin{figure}[t]
\centering
\includegraphics[width=8.5 cm]{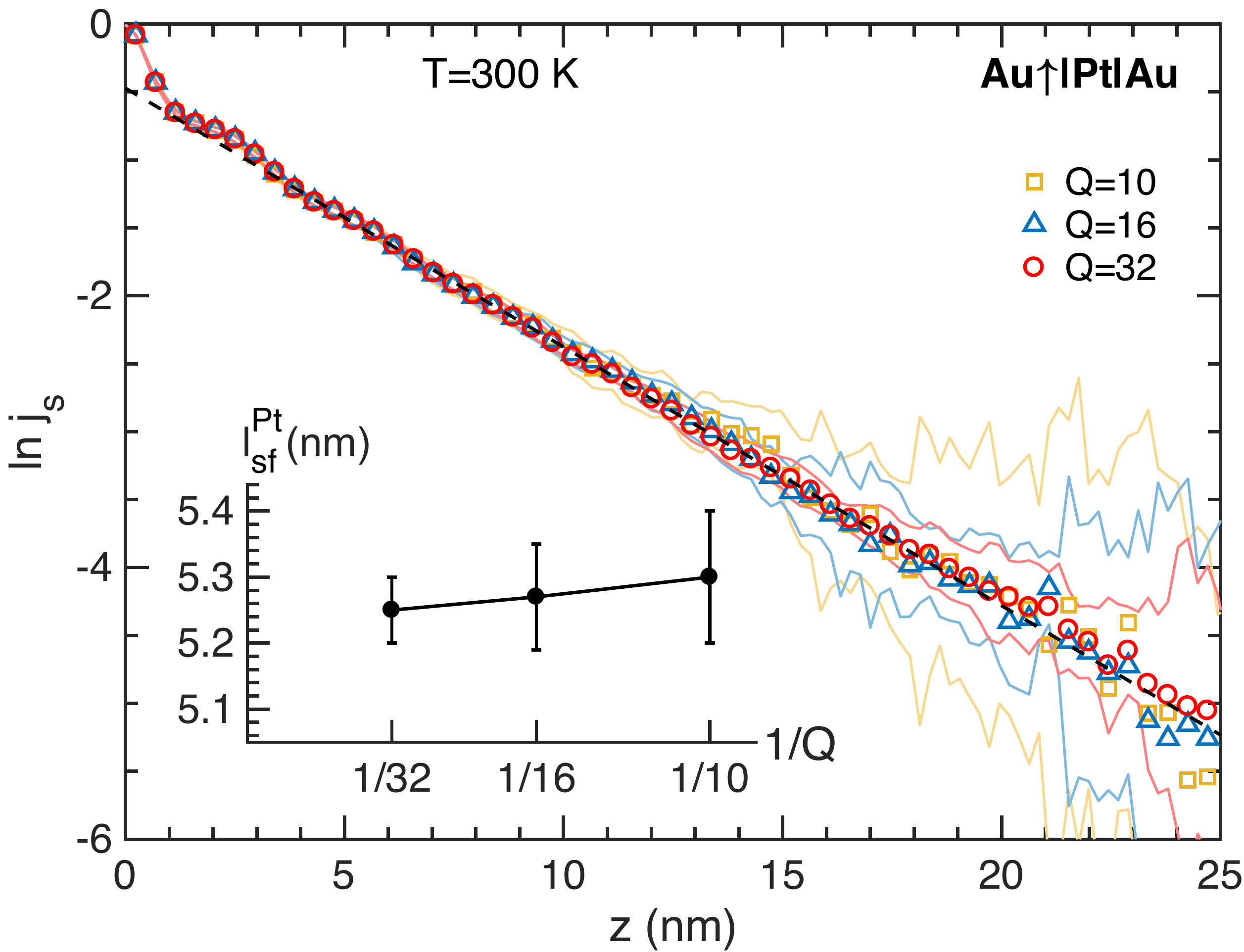}
\caption{Natural logarithm of the spin current density calculated for RT Pt with a 7$\times$7 lateral supercell and three different $Q \times Q$ samplings of the BZ: $Q=10$ (yellow squares), $Q=16$ (blue triangles) and $ Q=32$ (red circles). The color coordinated symbols and solid lines indicate the mean and a measure of the spread of the data, respectively, for 20 different configurations for different Q samplings. The dashed black line indicates the linear fit determined for $\ln j_{s}$ calculated with $Q=32$. Though the three curves initially overlap perfectly, for $z\ge 14 \,$nm we see that noise in the mean and spread of $\ln j_s$ for $Q=10$ is substantially larger than for the $Q=32$ data. Inset: $l_{\rm sf}^{\rm Pt}$ as a function of the BZ sampling parameter $Q$.}
\label{Fig7}
\end{figure}

The last point brings us to the question of BZ sampling. The spin current $j_s(z)$ is obtained by summing partial spin currents over a discrete grid of ${\bf k}_\parallel$ vectors in a 2D BZ and integrating over $xy$ planes of real space atomic layers. As the BZ grid becomes finer, the fluctuations in spin current density in each layer must tend towards a converged value dependent on the lateral supercell size. In Fig.~\ref{Fig7} we show the fluctuations found as a function of $z$ for a room temperature Pt slab of length $\sim 35\,$nm and an $N=7$ lateral supercell. We compare the results obtained for three $Q \times Q$ BZ sampling densities with $Q=10,16,32$. As the spin currents become smaller, the noise in the data becomes larger. The solid lines in Fig.~\ref{Fig7} are a measure of the spread found for 20 random configurations of disorder. The spread becomes significantly smaller with increasing $Q$. Since the current injected from the left lead is fully polarized, the noise does not significantly affect the determination of $l_{\rm sf}^{\rm Pt}$. We shall see in the next subsection that a smaller spin current entering from a diffuse Py$|$Pt interface leads to more noise in the data, making the choice of BZ sampling more critical.

\subsubsection*{Leads}
\label{subsec:leads}

\begin{figure}[t]
	\includegraphics[width=8.5 cm]{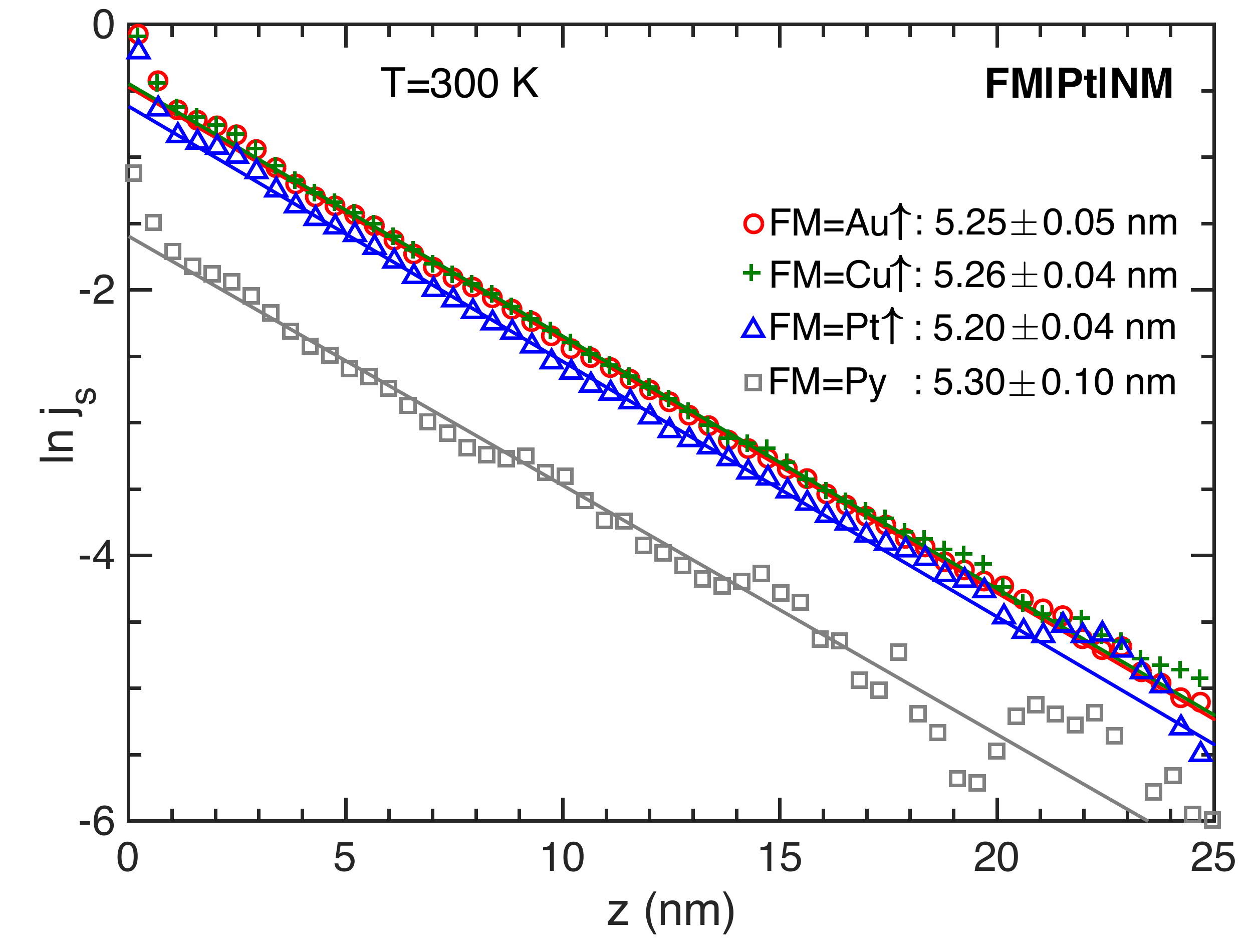} 
	\caption{Natural logarithm of the spin current injected into Pt as a function of the coordinate in the transport direction $z$ using different lead materials. The current was extracted from the results of a scattering calculation for a two-terminal NM$\uparrow$$|$Pt$|$NM$'$ configuration. The lattice constant of the Au leads was scaled to match to Pt. In the case of Cu, a lateral Cu $8 \times 8$ supercell was matched to a $2\sqrt{13} \times 2\sqrt{13}$ lateral Pt supercell. Injection from permalloy was realized in a Cu$|$Py$|$Pt$|$Cu configuration where lattice matching was realized using the same supercells as for Cu$|$Pt$|$Cu. The straight lines are weighted linear least squares fits from which the interface region is omitted; the error bars result from different ``reasonable'' weightings and cutoff values.}
	\label{Fig8}
\end{figure}

In Fig.~\ref{Fig1} we showed how $l_{\rm sf}$ obtained directly from the transmission matrix depended on the choice of lead material. Here we demonstrate that when determined from the decay of the spin current, $l_{\rm sf}$ does not depend on the lead material used. To study this, we carried out calculations for a 35 nm long slab of RT Pt with a $7 \times 7$ lateral supercell and a $32 \times 32$ BZ sampling for three different lead materials: ballistic HMF Cu$\uparrow$, Au$\uparrow$ and Pt$\uparrow$ leads, in each case raising the spin-down electronic bands above the Fermi energy by adding a constant to the AS potential. Thus, a fully polarized spin current enters Pt and decays exponentially as shown in Fig.~\ref{Fig8}. 

To within 1\%, $l_{\rm sf}^{\rm Pt}$ is the same for all lead materials. The quantum oscillations are also independent of the lead material supporting our assertion that they are an intrinsic property of Pt. What does change is the interface contribution to the loss of spin current (``spin memory loss'') as indicated by different intercepts for the three different HMF leads. 

Since these leads were polarized artificially, we also examine what happens when a ``naturally'' polarized current from a ferromagnetic material enters Pt. We used an $8 \times 8$ lateral supercell of Cu$|$Py to match to a $2\sqrt{13} \times 2\sqrt{13}$ lateral supercell of Pt and absorbed the residual mismatch in a small trigonal distortion of Pt. For this geometry, we find $l_{\rm sf}^{\rm Pt}=5.3\pm0.1~\rm nm$ in Fig.~\ref{Fig8} (grey squares). The slight difference from the other values can be traced to the small trigonal distortion of the Pt lattice. Compared to the HMF lead cases, a smaller spin current enters Pt from Py because (i) the current polarization in Py is not 100\% (see Sect.~\ref{SSec:beta_Py}) and (ii) because of the spin-flipping at the interface (spin-memory loss). As discussed in the previous subsection, the noise in $\ln j_s(z)$ for smaller absolute values of $j_s(z)$ at large $z$ could be reduced somewhat by increasing the BZ sampling.

\subsubsection*{Three center terms}
\label{subsecA:tct}
Including three center terms in the SOC part of the Hamiltonian increases the computational cost by $\sim 70\%$ \cite{Starikov:prb18}. The effect on $l_{\rm sf}^{\rm Pt}$ is compared for 5$\times$5 and 7$\times$7 supercells in Table \ref{tab:SCPt}. For a 5$\times$5 supercell we find that $l_{\mathrm Pt}$ decreases by 7.5\% from $5.65\pm0.08~\rm nm$ with two center terms to $5.22\pm0.09~\rm nm$ with three center terms. For a 7$\times$7 supercell, $l_{\rm sf}^{\rm Pt}$ decreases by 5.5\% from $5.25\pm0.05~\rm nm$ with two center terms to $4.96\pm0.07~\rm nm$ with three center terms. 

\subsubsection*{Basis: spd vs spdf}
\label{subsecA:basis}
Using a 16 orbital $spdf$ basis instead of a 9 orbital $spd$ basis increases the computational costs by a factor $(16/9)^3 \sim 5.6$. Thus, we use only a 5$\times$5 lateral supercell to estimate the effect of including $f$ orbitals on $l_{\rm sf}^{\rm Pt}$ compared with the $spd$ results in Table \ref{tab:SCPt}. We find a 7.5$\%$ decrease in $l_{\rm sf}^{\rm Pt}$ from $5.65\pm0.08~\rm nm$ with an $spd$ basis to $5.21\pm0.07~\rm nm$ with an $spdf$ basis.
In view of the substantial computational costs incurred in including them and their relatively small effect on $l_{\rm sf}$, neglect of the three centre terms in the Hamiltonian and $f$ orbitals in the basis is justified by the much larger uncertainty that currently exists in the experimental determination of $l_{\rm sf}$. The only barrier to including them, should the experimental situation warrant an improved estimate, is computational expense. \emph{Our best  estimate  of $l_{\rm sf}^{\rm Pt}$ at 300 K is $5.3\pm0.4$ nm. }

\subsection{${\boldsymbol \beta}$ for Permalloy}
\label{SSec:beta_Py}

\begin{figure}[t]
\centering
\includegraphics[width=8.5cm]{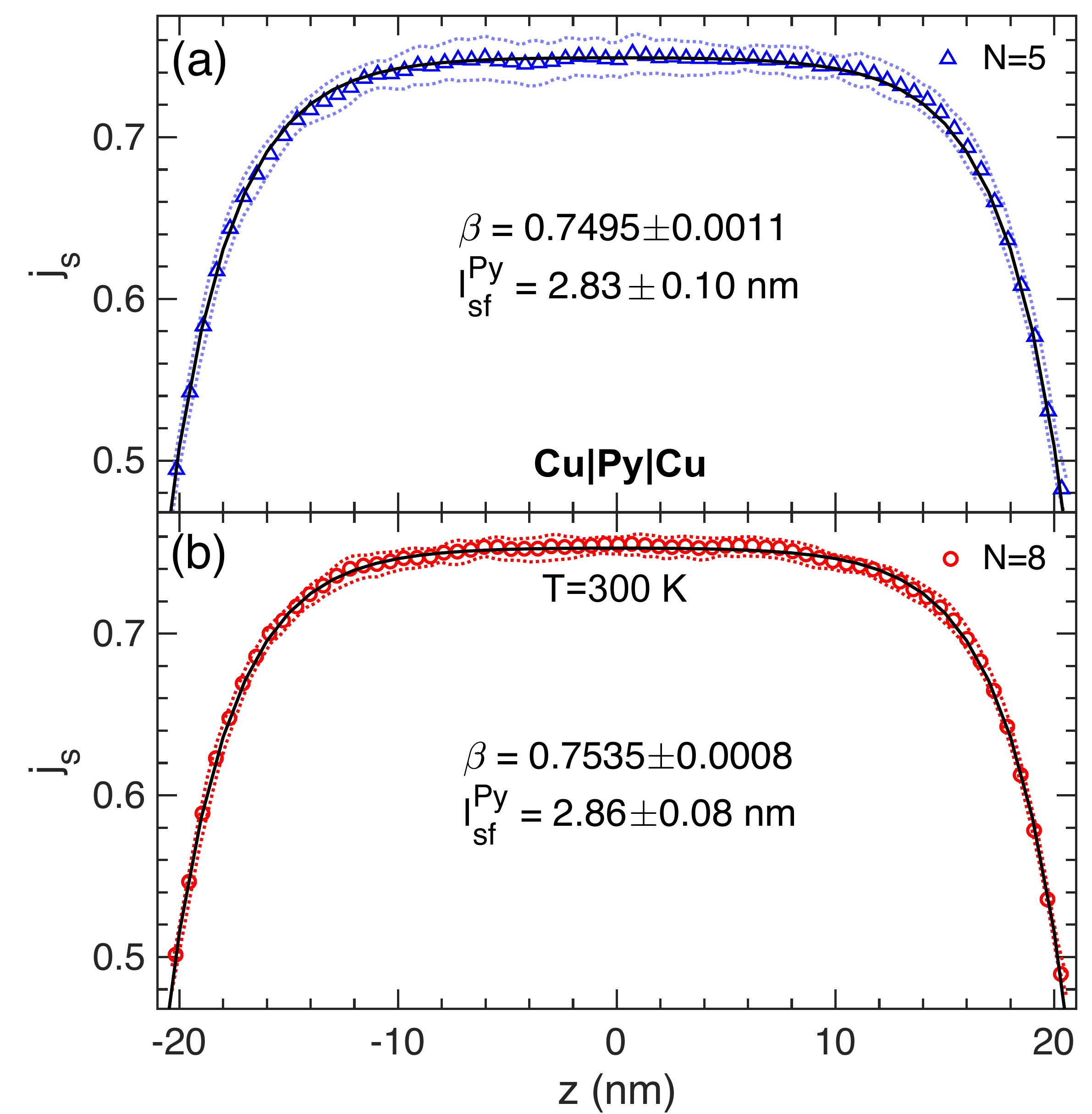}
\caption{A charge current passed through Py polarizes to $\beta$ in the middle of the scattering region. By fitting (solid black lines) the spin current calculated for an $N \times N$ supercell to Eq.~\eqref{eq:jss}, $\beta$ and $l_{\rm sf}^{\rm Py}$ are extracted for (a) $N=5$ (blue) and (b) $N=8$ (red). The dotted lines indicate the spread from 20 different configurations of disorder. }
\label{Fig9}
\end{figure}

To determine the transport polarization $\beta$ of Py, we can use a symmetric NM$|$FM$|$NM configuration and equation \eqref{eq:jss}. By choosing the center of the FM slab to be the origin $z=0$ with the interfaces at $z=\pm L/2$, $\widehat{j_s}(z) = \beta - c \cosh(z/l_{\rm sf})$ as in \eqref{eq:jss}. The results of injecting an unpolarized current from Cu leads into a 40~nm thick slab of RT Py are shown in Fig.~\ref{Fig9}(a) for a $5 \times 5$ lateral supercell. $\beta$ and $l_{\mathrm{sf}}^{\rm Py}$ are determined simultaneously by using both as free parameters for the fitting. 

Since the current polarization for an infinitely long Py slab should be $\beta$ for all $z$ and because $\cosh(0)=1$, $c$ must vanish in the limit $L\rightarrow\infty$ over a length scale $l_{\rm sf}$. Because the scattering region is finite in length, $\widehat{j_s}(z)$ must be fitted to $\beta - c(L)\cosh(z/l_{\rm sf})$. We need to determine $c(L)$ or ensure that it does not affect the values of $\beta$ and $l_{\rm sf}$ obtained by fitting. These values are given in Table~\ref{tab:SCPy} as a function of the length of the scattering region, $L_{\rm Py} \simeq 10,20,30,40$ nm. Reasonable estimates of $l_{\rm sf}$ and $\beta$ are found for $L \ge 20$ nm with very acceptable error bars. 

\begin{table}[b]
\caption{Dependence of the SDL and polarization $\beta$ on the length $L_{\rm Py}$ of the Py slab and on the $N \times N$ supercell size for Py at 300 K for $N=5,8$. Calculations are performed with k-point sampling equivalent to $140\times140$ for a $1\times1$ unit cell in each case.}
\begin{ruledtabular}
\begin{tabular}{rcccc}
$N$ & $L_{\rm Py}\,$(nm) 
            & \multicolumn{1}{c}{$l_{\rm sf}$ (nm)}  
                             & \multicolumn{1}{c}{$\beta$} \\
\hline
5   & 10.44 & $2.29\pm0.72$  &  $0.7200\pm0.0500$\\ 
    & 20.66 & $2.71\pm0.14$  &  $0.7410\pm0.0032$\\
    & 30.88 & $2.69\pm0.08$  &  $0.7481\pm0.0014$\\
    & 41.11 & $2.83\pm0.10$  &  $0.7495\pm0.0011$\\
\hline       
8   & 41.11 & $2.86 \pm 0.08$  &  $0.7535\pm0.0007$\\
\end{tabular}
\end{ruledtabular}
\label{tab:SCPy}  
\end{table}

\subsubsection*{Supercell size}
\label{subsecC:supercell}

We compare the results obtained for Py with 5$\times$5 and 8$\times$8 supercells in Fig.~\ref{Fig9} and Table~\ref{tab:SCPy}. Both thermal and chemical disorder contribute to the fluctuations which are larger for $N=5$ than for $N=8$. However, the parameter estimates do not show a significant $N$ dependence. Carrying out a calculation with $N=8$ would be necessary only in cases where statistical fluctuations or parameter errors are unacceptably large. 

\subsubsection*{Averaging L$\rightarrow$R and R$\rightarrow$L currents}

\begin{figure}[t]
\centering
\includegraphics[width=8.5cm]{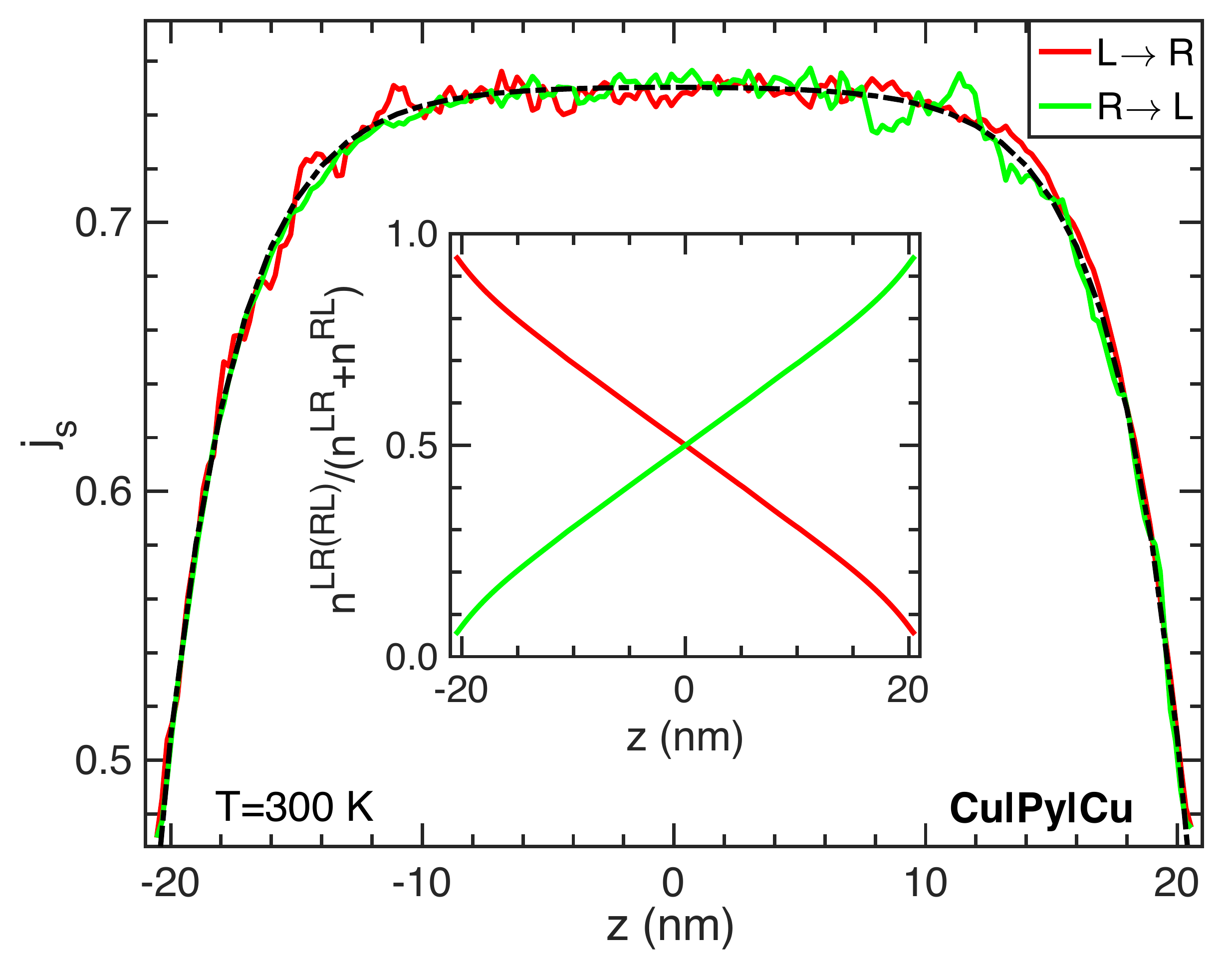}
\caption{Spin currents $\widehat{j}_{sz}(z)$ obtained by injecting electrons into Py from a left Cu lead (red) and from injecting holes from a right Cu lead (green) for an 5$\times$5 lateral supercell. The black dashed line shows the fit to the averaged current discussed in the text. Inset: Fractional nonequilibrium particle densities incident from left (red) and right (green) leads.}
\label{Fig10}
\end{figure}

In Fig. \ref{Fig10}, we plot the spin current $j_{sz}^{LR}(z)$ that arises when a current of electrons is passed from left to right, and $j_{sz}^{RL}(z)$ when a current of holes is passed from right to left, for a 40 nm long slab of an $5\times5$ supercell of Py sandwiched between ballistic Cu leads. Reflections at the Cu$|$Py interfaces give rise to interferences that slowly decay into the scattering region. The interference and its decay are clearly visible in Fig.~\ref{Fig10} for both currents as they progress from the source to the drain lead. The fluctuations are significantly reduced after averaging using \eqref{scrave}. 

\subsection{${\bf l_{\rm sf}}$ for Permalloy}
\label{SSec:lsf_Py}

\begin{figure}[t]
\includegraphics[width=8.5 cm]{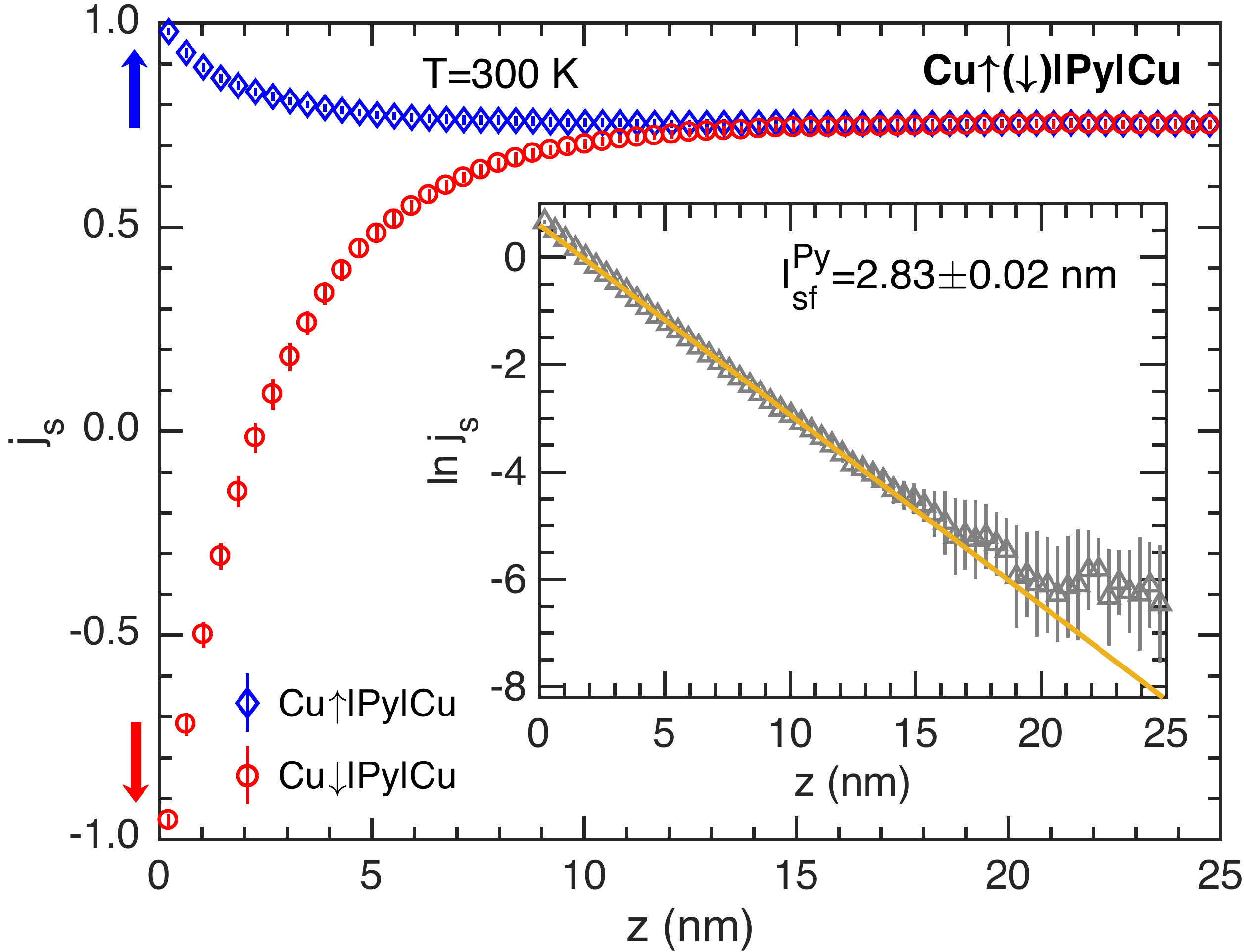} 
\caption{Fully polarized spin-up (blue) and spin-down (red) currents injected from the left ballistic Cu$\uparrow$, respectively Cu$\downarrow$ lead into 40~nm of Py with RT thermal lattice and spin disorder saturate to $\beta$ far from the lead. The difference of the two currents decays exponentially to zero. The supercell size is 8$\times$8 and the Brillouin zone k-sampling is 28$\times$28. Inset: natural log of the difference is fit linearly (in yellow) to yield $l_{\rm Py}=2.83\pm0.02$ nm.}
\label{Fig11}
\end{figure}

Although we obtain reasonable values for $l_{\rm sf}^{\rm Py}$ simultaneously with $\beta$, it can be desirable to be able to extract $l_{\rm sf}^{\rm Py}$ independently. For a symmetric NM$|$FM$|$NM configuration, the spin current has the form \eqref{eq:js} which approaches $\beta$ asymptotically for scattering regions much longer than $l_{\rm sf}$. Unlike $l_{\rm sf}^{\rm Pt}$ for which $\beta=0$, the finite asymptotic value of $\beta$ prevents us from extracting $l_{\rm sf}^{\rm Py}$ by taking the logarithm of $j_s(z)$. However, by considering NM$\uparrow$$|$FM$|$NM and NM$\downarrow$$|$FM$|$NM configurations for which $j_s^\uparrow(z) = 
\beta + b_\uparrow e^{-z/l_{\rm sf}} - a_\uparrow e^{z/l_{\rm sf}}$ and, respectively, $j_s^\downarrow(z) = 
\beta + b_\downarrow e^{-z/l_{\rm sf}}-a_\downarrow e^{z/l_{\rm sf}}$, we can take the difference so the constant $\beta$ drops out. We then consider a long scattering region and values of $z$ far from the right-hand interface so that the exponentially increasing terms can be neglected and we are left with a pure exponentially decreasing function. To optimize the ``systematic cancellation'' when taking the difference of the two spin currents, we use identical microscopic configurations of alloy and thermal disorder to perform scattering calculations first with Cu$\uparrow$ and then with Cu$\downarrow$ left leads. This is then done pairwise for 20 different configurations of 40~nm long disordered Py to obtain the results shown in Fig.~\ref{Fig11}. The small oscillations in the spin current found for Pt are not observed here for Py. Presumably, this is due to the larger amount of disorder, as we now also have thermal spin disorder and substitutional alloy disorder in addition to the thermal lattice disorder, resulting in a pure exponential decay of  $j_s^\uparrow(z)-j_s^\downarrow(z)$.

The natural logarithm of the difference is shown in the inset to Fig.~\ref{Fig11}. A weighted linear least squares fit yields a room temperature decay length of $l_{\rm sf}^{\rm Py} =  2.83 \pm 0.02$ nm for 8$\times$8 supercell. Only data between $z=2$ and 20~nm is used for the fitting. The small curvature around $z=0$ nm is due to spin-memory loss. Beyond $z=20$ nm the variance in the spin current is relatively larger, and the exponentially increasing term in $j_s^\uparrow(z)-j_s^\downarrow(z)$ is not negligible. Since the region of fitting is $\sim6 \, l_{\rm sf}^{\rm Py}$, these effects are of little consequence. The weights are selected to be the inverse of the variance of the spin currents due to different configurations. The error bar is then estimated using weighted residuals. It is worth emphasizing that the value of $l_{\rm sf}^{\rm Py}=2.83\pm0.02$~nm obtained using HMF leads is in perfect agreement with the value $l_{\rm sf}^{\rm Py}=2.86\pm0.08$ (Fig.~\ref{Fig9} and Table~\ref{tab:SCPy}) obtained by passing a current from unpolarized NM leads. \emph{For Py at room temperature, our best estimate of $l_{\rm sf}^{\rm Py}$ is $2.8\pm0.1$~nm and of $\beta$ is $0.75\pm0.01$.}

\subsection{Spin Hall angle for Platinum}
\label{SSec:sha_Pt}

\begin{figure}[t]
\centering
\includegraphics[width=8.5 cm]{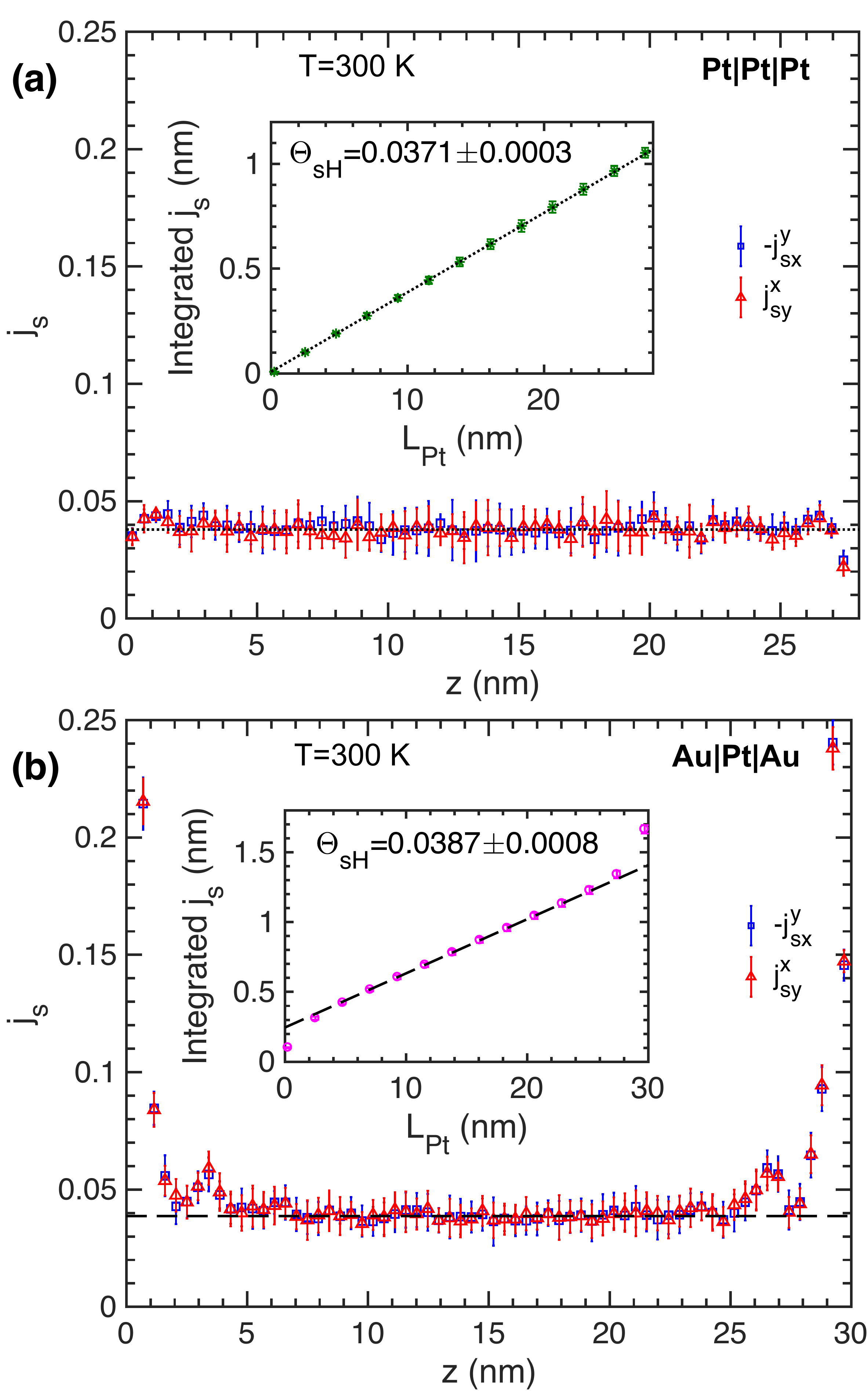}
\caption{Transverse spin currents driven by a charge current in the $z$ direction for a (111) oriented Pt slab embedded between (a) Pt and (b) Au leads. The error bars are the mean deviation of the currents for 20 configurations of disorder. The horizontal dotted and dashed lines indicate the extracted values of $\Theta_{\rm sH}$. For both leads, 7$\times$7 supercells were used with a 22$\times$22 BZ sampling. Inset: Integrated transverse spin currents as a function of $L_{\rm Pt}$ for a RT diffusive Pt scattering region embedded between ballistic Pt (green stars) (a) and Au (pink circles)  (b) leads. The dotted and dashed lines indicate the weighted linear least squares fit with Pt (a) and Au (b) leads, respectively. The interface contributions in (a) are negligible compared to (b).
}
\label{Fig12}
\end{figure}

A charge current passed through a length of diffusive Pt sandwiched between ballistic Pt leads results in spin currents in the transverse direction; this is the spin Hall effect \cite{Dyakonov:pla71, Hirsch:prl99, Zhang:prl00, Hoffmann:ieeem13, Sinova:rmp15}. The polarization direction of the spin current is given by a vector product of the original current direction ($z$) and the transverse spin current direction. Thus, spin currents in the $x$ and $-y$ directions are polarized in the $y$ and $x$ directions, respectively and have the same amplitude reflecting the axial symmetry of the system. These two transverse currents normalized to the longitudinal charge current, $\widehat{j}_{sy}^x(z)$ and $-\widehat{j}_{sx}^y(z)$, are plotted in Fig.~\ref{Fig12}(a) for a RT (111) oriented slab of Pt. The fluctuations about the bulk value are a result of a combination of configuration averaging, supercell size and BZ sampling. 

We extract the bulk value of the spin-Hall angle $\Theta_{\rm sH}$ as follows. Starting from the left interface at $z=0$, the configuration average of $\widehat{j}_{sy}^x(z)$ and $\widehat{j}_{sx}^y(z)$ is integrated over atomic layers up to some  $L_\mathrm{Pt}$: $J_s^\perp(L_{\rm Pt})=\int_0^{L_\mathrm{Pt}}[\, \widehat{j}_{sy}^x(z)-\widehat{j}_{sx}^y(z)]/2~dz$. The integrated quantities for a number of discrete values of $L_\mathrm{Pt}$ are shown in the inset to Fig.~\ref{Fig12}(a) as green stars. A least squares fit to linear behaviour yields a value of $\Theta_{\rm sH}=3.71 \pm 0.03 \%$ as the slope \cite{WangL:prl16}. The error bar results from the weighted residuals where the weights are the mean deviation for 20 configurations of thermal disorder. 

The above calculations were carried out with a 7$\times$7 lateral supercell and a 22$\times$22 BZ sampling that is equivalent to a 154$\times$154 sampling for a 1$\times$ 1 unit cell. We now examine the effect on $\Theta_{\rm sH}$ of varying some of the different computational parameters discussed in Sec.~\ref{SSec:scatcalc}.

\subsubsection*{Leads}
\label{subsecD:lead}

To rule out an eventual dependence of $\Theta_{\rm sH}$ on the leads, results for Pt and Au leads are compared in Figs.~\ref{Fig12}(a) and (b) respectively. Close to the Au leads, the transverse spin currents are dominated by a huge Au$|$Pt interface contribution \cite{WangL:prl16} and then drop rapidly towards the bulk value, indicated by the horizontal dashed line, away from the two interfaces. The interface contributions with Pt leads are negligible compared to Au. The slopes determined by linear least squares fitting of the integrated spin current density are nearly identical. To ensure a sufficiently long range in $L_{\rm Pt}$ that exhibits linear behaviour, a longer length of Pt must be used with Au leads than with Pt leads. 

\subsubsection*{Supercell size and k-point sampling}
\label{subsecD:supercell}

\begin{table}[b]
\caption{Dependence of $\Theta_{\rm sH}$ (in \%) for RT Pt on the supercell size $N$, without and with three centre SOC terms. Calculations were performed with a $Q \times Q$ k-point sampling nearly equivalent to $160\times160$ for a $1\times1$ supercell in each case; $N \times Q \sim$160.}
\begin{ruledtabular}
\begin{tabular}{rccrcc}
    &     & & \multicolumn{2}{c}{$spd$}            & \multicolumn{1}{c} {$spdf$} \\
\cline{4-5}\cline{6-6} 
$N$ & $Q$ & $N \times Q$      
                & \multicolumn{1}{c} {2 center}
                                   & \multicolumn{1}{c} {3 center}
                                                   & \multicolumn{1}{c} {2 center} \\
\hline
 3  &  54 & 162 & $3.65 \pm 0.07$  &               &                 \\
 5  &  32 & 160 & $3.79 \pm 0.06$  & $5.1 \pm 0.2$ & $2.95 \pm 0.03$ \\    
 7  &  22 & 154 & $3.71 \pm 0.03$  & $5.0 \pm 0.1$ &                 \\      
 7  &  32 & 224 & $3.73 \pm 0.03$  &               &                 \\ 
10  &  16 & 160 & $3.75 \pm 0.01$  &               &					    
\end{tabular}
\end{ruledtabular}
\label{tab:thetasc}  
\end{table}

We studied how the SHA depends on the size of the lateral supercell with $N=3,5,7,10$ using a BZ sampling $Q$ for each $N$ that corresponds to sampling a 1$\times$1 unit cell with $NQ \sim 160$ k points. Unlike $l_{\rm sf}^{\rm Pt}, \Theta_{\rm sH}$ shows a negligible dependence on the supercell size, as seen in Table~\ref{tab:thetasc} for results calculated with two center SOC terms. On changing $N$, the central value scarcely changes with respect to the value $\Theta_{\rm sH}=3.71\%$ found above. What does change is that the already small error bar  decreases with increasing supercell size.

Calculating $\Theta_{\rm sH}$ for a 7$\times$7 Pt supercell with a denser k-sampling, $Q=32$ ($NQ=224$), yields $\Theta_{\rm sH}=3.73\%$ compared to $\Theta_{\rm sH}=3.71\%$ with $Q=22$ ($NQ=154$), see Table~\ref{tab:thetasc}. Thus  a choice of $Q=22$ for $N=7$ is quite sufficient.

\subsubsection*{SOC: three center terms}
\label{subsecD:tct}

The results obtained with the three center terms in the SOC Hamiltonian included are also given in Table \ref{tab:thetasc}. $\Theta_{\rm sH}$ increases by about a third compared to the values with two center terms. Three center (3C) terms are thus seen to affect $\Theta_{\rm sH}$ much more than $l_{\rm sf}^{\rm Pt}$. We return to this below.

\subsubsection*{Basis: spd vs spdf}
\label{subsecD:basis}

\begin{figure}[b]
\centering
\includegraphics[width=8.4 cm]{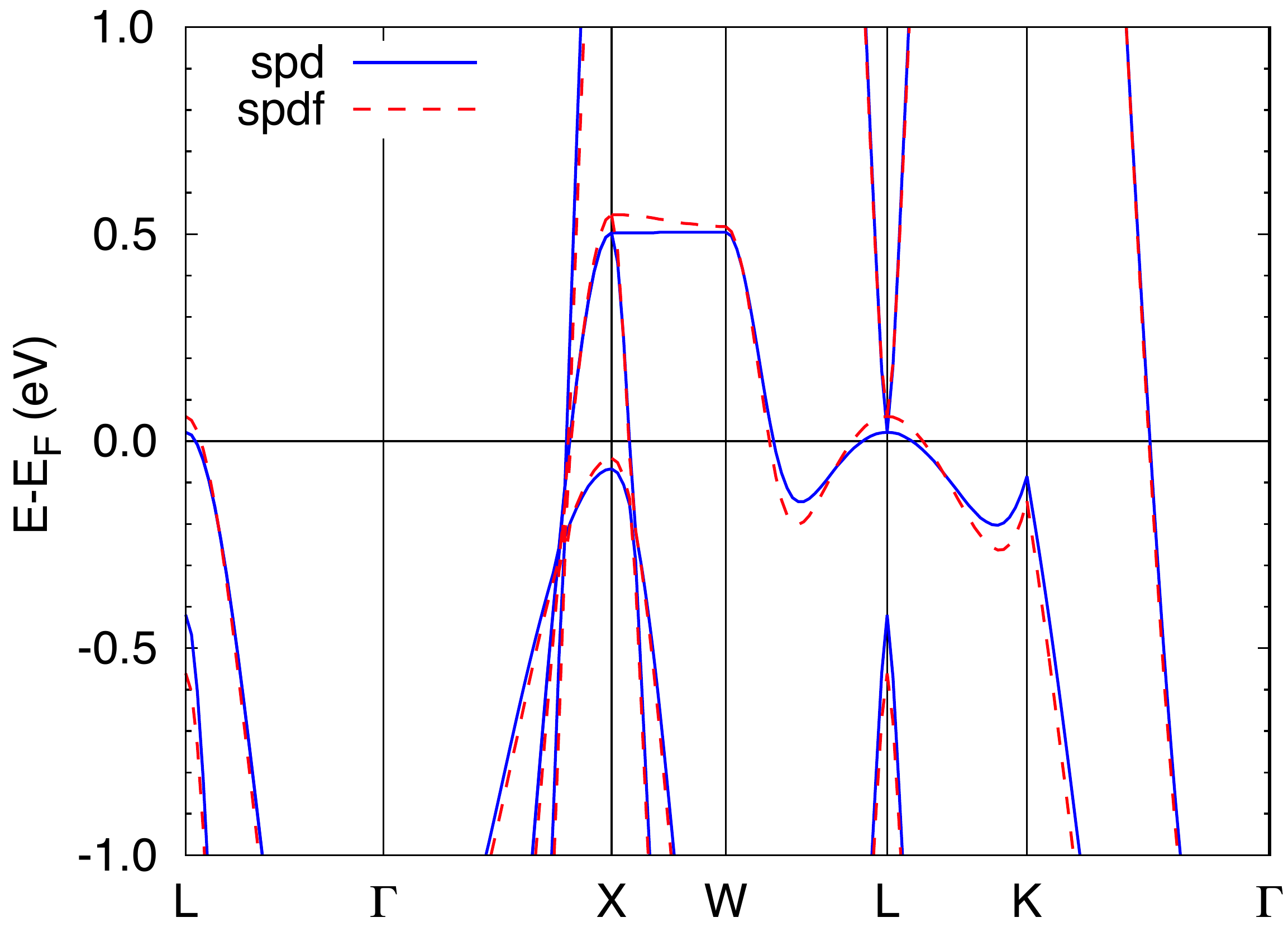}
\caption{Band structure of Pt evaluated using Stuttgart LMTO code with an $spd$ (blue) and $spdf$ (red) basis. The dispersion about the Fermi energy and details of the Fermi surface are very sensitive to the choice of basis.}
\label{Fig13}
\end{figure}

Augmenting the $spd$ basis with $f$ orbitals increases the computational effort by a factor $(16/9)^3 \sim 5.6$ and reduces $\Theta_{\rm sH}$ by about a fifth from $3.71\%$ to $2.95\%$. The sensitivity of the Pt spin-Hall angle to the basis and three-center terms can be related \cite{Guo:prl08} to the sharp peak in the density of states (DoS) at the Fermi energy, $D(E_F)$, that originates in the very flat X-W-L-K $d$ band \cite{Andersen:prb70} whose dispersion depends sensitively on the choice of basis, \cref{Fig13}. The spin-orbit splitting of the unoccupied orbitally doubly degenerate X-point state at $\sim 0.5$~eV is 0.66~eV and of the unoccupied L-point state just above the Fermi energy is even larger at 0.93~eV \cite{Guo:prl08}. These splittings are so large that the effect of not recalculating the Fermi energy when SOC is included needs to be examined. 

\subsubsection*{SOC self-consistency}

So far we have determined $l_{\rm sf}$ and $\Theta_{\rm sH}$ for Pt using AS potentials calculated self-consistently with the Stuttgart TB-LMTO code  for $spd$ and $spdf$ bases without (w/o) SOC; in the latter, the $f$ states were included by downfolding. For the scattering calculations, these potentials were used to construct a Hamiltonian matrix including the spin-orbit interaction $H_{\rm so}$ in an LMTO basis \cite{Starikov:prb18} but using the Fermi energies calculated without SOC. The results obtained with an $spd$ or $spdf$ basis using only two center terms in $H_{\rm so}$ are reproduced in the first row of Table \ref{tab:potPt} from \cref{tab:SCPt} ($l_{\rm sf}$) and \cref{tab:thetasc} ($\Theta_{\rm sH}$) and labelled ``w/o SOC''. 

To include SOC self-consistently, we generate new $spd$ and $spdf$ potentials for Pt as input for scattering calculations using a version of the Stuttgart LMTO-ASA code extended to include SOC self-consistently \cite{footnote4}. The results obtained with these potentials, labelled ``with SOC'' are shown in the second row of Table \ref{tab:potPt}. The change found in Sec.~\ref{subsecA:basis}  for $l_{\rm sf}^{\rm Pt}$ on going from an $spd$ to an $spdf$ basis is almost completely eliminated for the self-consistent SOC potentials to yield a best estimate of $l_{\rm sf}^{\rm Pt} =  5.3\pm0.4$~nm. 

For $\Theta_{\rm sH}$, the discrepancy between values found with $spd$ and $spdf$ bases remains. With an $spdf$ basis we find $\Theta_{\rm sH} = 3.16 \pm 0.02\%$. Including a correction for three center terms of $5.0 - 3.7\%$ from \cref{tab:thetasc}, {\em our best estimate for $\Theta_{\rm sH}^{\rm Pt}$ is $3.2 + 5.0-3.7= 4.5 \%$ with an uncertainty of about one percent}.

\begin{table}[h]
\caption{Dependence of the SDL $l_{\rm sf}$ and the spin Hall angle $\Theta_{\rm sH}$ for RT Pt on whether or not the Fermi energy was calculated without (w/o) or with SOC. To compare the results obtained with $spd$ and $spdf$ bases, only two-center terms in the SOC were included in the scattering calculations. A 5$\times$5 supercell was used with a k-point sampling equivalent to $160\times160$ for a $1\times1$ supercell.}
\begin{ruledtabular}
\begin{tabular}{ccccc}
         & \multicolumn{2}{c}{$l_{\rm sf}$ (nm)}
                            & \multicolumn{2}{c}{$\Theta_{\rm sH}(\%)$} \\
\cline{2-3} \cline{4-5}
     & \multicolumn{1}{c}{$spd$} 
                            & \multicolumn{1}{c}{$spdf$} 
                                              & \multicolumn{1}{c}{$spd$} 
                                                                & \multicolumn{1}{c}{$spdf$} \\
\hline
w/o SOC  & $5.65 \pm 0.08$  & $5.21 \pm 0.07$ & $3.79 \pm 0.06$ & $2.95\pm0.03$\\
with SOC & $5.28 \pm 0.09$  & $5.30 \pm 0.09$ & $4.27 \pm 0.03$ & $3.16\pm0.02$\\    
\end{tabular}
\end{ruledtabular}
\label{tab:potPt}  
\end{table}

\section{Comparison with other work}
\label{Sec:Comp}

\begin{table*}[t]
\caption{Experimental values of room temperature spin-flip diffusion length $l_{\rm sf}^{\rm Pt}$ and spin Hall angle $\Theta_{\rm sH}^{\rm Pt}$ for Pt. These are divided into work that took interface SML or transparency into account in their analysis (lower) and work that did not (upper). The bottom line contains our best theoretical estimates calculated using disorder that reproduces the experimental room temperature resistivity. SHE-STT-FMR: spin Hall effect spin-transfer torque ferromagnetic resonance. SP-ISHE: spin pumping - Inverse Spin Hall effect. SHM: spin Hall magnetoresistance. NL-SA-ISHE: nonlocal spin absorption - ISHE. HR: Harmonic Response. MOKE: Magneto-optical Kerr effect. VNA-FMR: vector network analyser FMR.
}
\begin{ruledtabular}
\begin{tabular}{llllll}
\multicolumn{1}{c}
{$\rho^{\rm Pt} (\mu \Omega\,$cm)}  
             & \multicolumn{1}{c}{$l_{\rm sf}^{\rm Pt}$(nm)}
                            & \multicolumn{1}{c}{$\rho l_{\rm sf}({\rm f}\Omega {\rm m}^2)$}
                                             &  $\Theta_{\rm sH}^{\rm Pt}$ (\%) 
                                                           & Method      & Reference \\
\hline
42 \cite{Mosendz:prl10}  
             & $3.7\pm0.2$ 	& 1.55 & $8   \pm 1   $ & SP-ISHE     &   Azevedo PRB11 \cite{Azevedo:prb11} \\
20 	         & $1.4\pm0.3$ 	& 0.28 & $6.8 \pm 0.5 $ & SHE-STT-FMR &     Liu arXiv11 \cite{Liu:arxiv11} \\
$23\pm1$     & $8.3\pm0.9$  & 1.9  & $1.2 \pm 0.2 $ & SP-ISHE     &      Feng PRB12 \cite{Feng:prb12} \\
28           & $1.2\pm0.06$ & 0.34 & $2.2 \pm 0.4 $ & SHE-STT-FMR &    Kondou APE12 \cite{Kondou:ape12} \\
--           & $7.7\pm0.7$  & --   & $1.3 \pm 0.1 $ & SP-ISHE     &  Nakayama PRB12 \cite{Nakayama:prb12} \\
--           & $1.5\pm0.5$  & --   & $11  \pm 8   $ & SHM         & Althammer PRB13 \cite{Althammer:prb13} \\
--           & 1.2          & --   & $8.6 \pm 0.5 $ & SP-ISHE     &     Zhang APL13 \cite{ZhangW:apl13} \\
48           & 7.3          & 3.5  & $10  \pm 1   $ & SP-ISHE     &      Wang PRL14 \cite{Wang:prl14} \\
28			 & $2.1\pm0.2$	& 0.59 & $2.2 \pm 0.8 $ & SHE-STT-FMR &   Ganguly APL14 \cite{Ganguly:apl14} \\
39.7         & $2.0\pm2.2$  & 0.79 & $1.5 \pm 2.9 $ & NL-SA-ISHE  &     Isasa PRB15 \cite{Isasa:prb15a, *Isasa:prb15b} \\
10.12        & $6.5\pm0.1$  & 0.66 & $2.2 \pm0.3  $ & NL-SA-ISHE  &   Sagasta PRB16 \cite{Sagasta:prb16} \\
\hline
$17.9\pm0.2$ & $3.4\pm0.4$  & 0.61 & $5.6 \pm 1.0 $ & SP-ISHE     & 
                                                   Rojas-S{\'{a}}nchez PRL14 \cite{Rojas-Sanchez:prl14} \\
$15\pm1$     & $1.4\pm0.2$  & 0.21 & $19  \pm 4   $ & SHE-STT-FMR &    Zhang NatM15 \cite{ZhangW:natp15} \\
15           & $5.1\pm0.5$  & 0.77 & $8.9 \pm 0.3 $ & HR          &    Nguyen PRL16 \cite{Nguyen:prl16} \\
20.6         & $11\pm3$     & 2.27 & $8   \pm 2   $ & MOKE        &     Stamm PRL17 \cite{Stamm:prl17} \\
18.8-21.3    & $8.0\pm0.5$  & 1.60 & $3.0 \pm 0.2 $ & SP-ISHE     &        Tao SA18 \cite{Tao:sca18} \\
16.3         & $4.2\pm0.1$  & 0.68 & $38.7\pm 0.8 $ & VNA-FMR     &    Berger PRB18 \cite{Berger:prb18b} \\
\hline
$10.8\pm0.2$ & $5.3\pm0.4$  & 0.57 & $4.5 \pm 1   $ & Ab-initio   & This work \\
\end{tabular}
\end{ruledtabular}
\label{tab:Pt_expt}  
\end{table*}

\subsection*{Experiment}

A 2007 review \cite{Bass:jpcm07} of spin-diffusion lengths in metals and alloys contains a single entry for Pt (also for Nb, Pd, Ru, and W) and just a handful for Py. The entry for Pt refers to measurements at low temperatures necessitated by the use of superconducting leads in conjunction with spin-valves (SV) in a CPP geometry \cite{Kurt:apl02}. These SV measurements were interpreted within the framework of diffusive transport and led to an estimate of ${l_{\rm sf}^{\rm Pt}\sim 14 \pm 6 \,}$nm but without a clear picture as to the microscopic origin of the diffusive scattering at the liquid He measurement temperatures. A common refrain in this section will be the need for detailed characterization of samples relating their transport properties to their microscopic structures and composition in order to make further progress. 

\subsubsection*{Pt: $l_{\rm sf}$ and $\Theta_{\rm sH}$}

At about the same time, the first electrical measurement of an ISHE was reported  for the light metal Al \cite{Valenzuela:nat06}. Although the nonlocal measurement technique used was not directly applicable to heavy metals like Pt with short SDLs \cite{Kurt:apl02}, it did herald the development of a number of new methods that were potentially suitable \cite{Hoffmann:ieeem13, Sinova:rmp15}. The first spin-pumping (SP-ISHE) \cite{Saitoh:apl06, Ando:prl08, Mosendz:prl10, *Mosendz:prb10}, nonlocal spin-absorption (NL-SA) \cite{Kimura:prl07, Vila:prl07}, spin-transfer torque FMR (SHE-STT-FMR) \cite{Liu:prl11} measurements established the feasibility of measuring the I(SHE) for materials like Pt but quantitative estimates of $\Theta_{\rm sH}$ required knowledge of $l_{\rm sf}$; extensive use was made of the only value available at the time from the low temperature CPP-SV measurements \cite{Kurt:apl02}.
Some ten years later, a review contained 22 room temperature entries for Pt \cite{Sinova:rmp15} with $l_{\rm sf}$ ranging from $1.2\pm 0.1$ to $11 \pm 2 \,$nm and $\Theta_{\rm sH}$ from $0.37$ to $12 \pm 4 \%$. We briefly discuss (some of) these experimental determinations in order to identify what needs to be done to improve the confrontation of theory and experiment.

Even for groups performing the same measurements, large differences emerged. Mosendz et al. \cite{Mosendz:prb10} reported $\Theta_{\rm sH}=1.3 \pm 0.2 \%$ using a value of $l_{\rm sf}=10 \pm 2 \,$nm they (incorrectly?) attributed  to Kurt et al. \cite{Kurt:apl02} Performing essentially the same SP-ISHE measurements, Azevedo et al. \cite{Azevedo:prb11} were able to determine a value of $l_{\rm sf}=3.7 \pm 0.2 \,$nm by varying the thickness of Pt that then yielded an estimate for $\Theta_{\rm sH}=8 \pm 1 \%$. However, they used as input the Pt conductivity measured by Mosendz et al. \cite{Mosendz:prl10} though such properties are very sensitive to where and how samples are prepared. Because of such sample to sample variability, it is very desirable to measure as many properties as possible on the same samples.

\begin{table*}[t]
\caption{Experimental values of room temperature spin-flip diffusion length $l_{\rm sf}^{\rm Py}$ and spin polarization $\beta$ for Py compared with our best estimated values values calculated with lattice and spin disorder that reproduce the experimental room temperature resistivity and magnetization. NLSV: nonlocal spin valve. SSE-ISHE: spin Seebeck effect + inverse spin Hall effect. SA-LSV: spin absorption in lateral spin valves. SW-DS: spin wave Doppler shift.}
\begin{ruledtabular}
\begin{tabular}{cllll}
\multicolumn{1}{c}
{$\rho (\mu \Omega\,$cm)}  
         & \multicolumn{1}{c}{$l_{\rm sf}^{\rm Py}$(nm)}
                             & \multicolumn{1}{c}{$\beta$}
                                                & \multicolumn{1}{c}{Method}  
                                                            & \multicolumn{1}{c}{Ref.} \\
				\hline
26.8 	     & 3 	 		 & 0.25 		    & NLSV  	& Kimura PRB05 \cite{Kimura:prb05} \\
23.1         & 4.5   		 & 0.49 		    & NLSV 		& Kimura PRL08 \cite{Kimura:prl08} \\
$\sim$30     & 2.5  		 & - 			    & SSE-ISHE  & Miao PRL13 \cite{Miao:prl13} \\
44           & $2.30\pm0.61$ & $0.31\pm0.02$    & SA-LSV 	& Sagasta APL17 \cite{Sagasta:apl17} \\
$29\pm3$     & -			 & $0.61\pm0.02$    & SW-DS 	& Zhu PRB10 \cite{Zhu:prb10} \\
25		     & -			 & 0.71			    & SW-DS 	& Haidar PRB13	\cite{Haidar:prb13} \\
\hline
$15.4\pm0.2$ & $2.8\pm0.1$ & $0.75\pm0.01$ & Ab-initio & This work \\
\end{tabular}
\end{ruledtabular}
\label{tab:lsf_Py_expt}  
\end{table*}

In the work cited in \cref{tab:Pt_expt}, both $l_{\rm sf}$ and $\Theta_{\rm sH}$ were extracted from measurements on the same samples, usually by varying the thickness of the Pt layer.
In the experimental results shown in the top half of the table, no attempt was made to take the interface properties of the FM$|$Pt or NM$|$Pt interfaces into account and we see that $l_{\rm sf}^{\rm Pt}$ ranges between 1.2 and 8.3 nm, while $\Theta_{\rm sH}$ lies in the range 1-11\%. 
The realization that interfaces play an essential role in degrading spin currents \cite{Rojas-Sanchez:prl14, LiuY:prl14, Nguyen:jmmm14} and that $l_{\rm sf}$ might be correlated with the enhancement of thin film resistivities by interface and surface scattering \cite{Nguyen:prl16} seemed to offer the possibility to resolve the difficulty posed by the spread in $l_{\rm sf}$ and $\Theta_{\rm sH}$ values. 

However, if we look at the work cited in the bottom half of \cref{tab:Pt_expt} that attempted to take interface SML or transparency into account, the situation has if anything worsened. We see that $l_{\rm sf}^{\rm Pt}$ ranges from 1.4 to 11 nm and find values for $\Theta_{\rm sH}$ as low as 3\% and as high as 39\%. Whereas the RT resistivity of bulk crystalline Pt is known to be 10.8 $\mu\Omega$~cm \cite{HCP90}, we see a wide range of resistivities, from 20 to 48 $\mu\Omega$~cm. Because it has long been known that the scattering from surfaces and interfaces in thin films leads to enhanced resistivity this is not very surprising. However, little is known about the microscopic nature of the corresponding disorder on an atomic scale making it difficult to predict how it might affect the SDL and spin Hall effect. The product $\rho l_{\rm sf}$ is seen to span a much larger range between 0.21 and 2.27 ${\rm f}\Omega {\rm m}^2$. In view of the values of $l_{\rm sf}^{\rm Pt}$ and $\Theta_{\rm sH}$ that we calculate for bulk Pt, we can only conclude that many experiments are not at present probing the corresponding phenomena in bulk materials but are dominated by extrinsic effects -- a situation very reminiscent of the discussion relating to the polarization of ferromagnets until the current-induced spin wave Doppler measurement technique was developed capable of probing the polarization far from surfaces and interfaces \cite{Vlaminck:sc08}. 


\subsubsection*{Py: ${l_{\rm sf}}$ and $\beta$}
 
Though it is used in a wide range of experiments and much is known about its magnetic properties, relatively few studies have been made of the transport parameters of bulk Permalloy at room temperature. These are compiled in Table \ref{tab:lsf_Py_expt} together with our best RT estimates of $l_{\rm sf}^{\rm Py}=2.8$ nm and $\beta=0.75$. With the exception of Kimura's 2008 value \cite{Kimura:prl08} and in spite of the reported resistivities being much higher than the bulk value of $\rho_{\rm Py}=15.4 \, \mu\Omega\,$cm \cite{Ho:jpcrd83}, there is excellent agreement between values of $l_{\rm sf}^{\rm Py}$ extracted from various experiments \cite{Kimura:prb05, Miao:prl13, Sagasta:apl17} and our best theoretical estimte. The polarizations reported from the non-local spin valve experiments \cite{Kimura:prb05, Kimura:prl08, Sagasta:apl17} are however much smaller than our bulk value, $\beta=0.75\pm0.01$. Two studies \cite{Zhu:prb10, Haidar:prb13} measured $\beta$ independent of $l_{\rm sf}$ using spin-wave Doppler shift experiments. Haidar and Bailleul \cite{Haidar:prb13} carried out systematic thickness dependent measurements at room temperature and predicted an extrapolated bulk value of 0.71. In the non-local spin valve based spin absorption experiments where $l_{\rm sf}^{\rm Py}$ and $\beta$ were determined simultaneously, the assumption of transparent Py$|$Cu interfaces may have affected the determination of $\beta$ \cite{Kimura:prb05, Kimura:prl08, Sagasta:apl17}. Alternatively, with \cref{Fig9} in mind, it is tempting to speculate that these experiments are probing an interface property rather than a property of bulk Py.  

\subsection*{Other calculations}

We are not aware of any theoretical studies of $l_{\rm sf}$ in either Pt or Py. There have been a number of studies of the ``intrinsic'' spin Hall conductivity (SHC) of bulk Pt that only depends on the electronic band structure of the crystalline material and can be evaluated in linear response by taking the $\omega \rightarrow 0$ limit of the optical conductivity using electronic structures calculated from first principles \cite{Guo:prl08} or tight binding fits to first principles band structures \cite{Tanaka:prb08}. In materials with strong spin-orbit coupled bands, it would appear that the intrinsic contribution dominates the SHC. The largest contributions arise from orbital degeneracies close to the Fermi energy. The inclusion of finite temperatures for the electrons via the Fermi Dirac function leads to a rapid quenching of the SHC in this picture \cite{Guo:prl08}.

\begin{figure}[t]
\centering
\includegraphics[width=8.4cm]{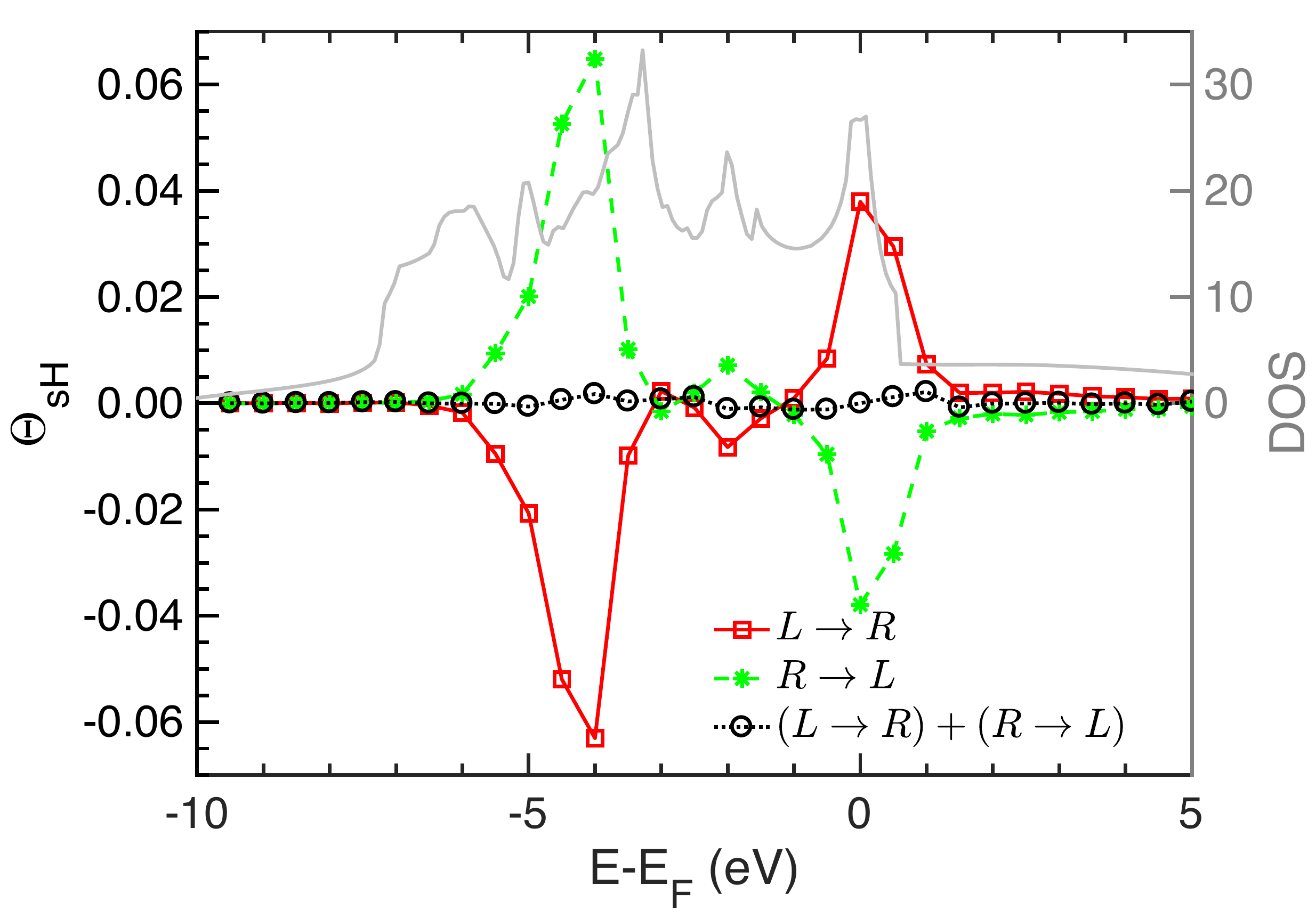}
\caption{Spin Hall angle as a function of the energy in a rigid band approximation  calculated for electrons incident from the left (red squares) and from the right (green stars). The sum of these contributions is shown as black circles. For reference, we show the Pt $d$-density of states in grey.}
\label{Fig14}
\end{figure}

Our calculation of the SHA is also ``intrinsic'' in the sense that no impurities are involved but as shown by Wang et al. it also leads to a different temperature dependence. In Ref.~\onlinecite{WangL:prl16} we found that the SHA is essentially linear in temperature so that the SHC is temperature independent. In \cref{Fig14}, we show how the SHA depends on the band filling. Like Guo we identify two prominent peaks that arise when the Fermi level coincides with orbital degeneracies at high symmetry points in the Brillouin zone. These features quite clearly survive lattice disorder. By calculating the contribution to the SHA from electrons propagating from $L \rightarrow R$ and from $R \rightarrow L$, we can obtain the so called ``Fermi sea'' contribution to the SHA by direct summation. Unlike the case of charge transport where filled bands make no contribution, there is no guarantee that this will always be the case for spin transport \cite{Lowitzer:prl11, Turek:prb14}.
 In the absence of disorder, time reversal and inversion symmetry lead to Kramers degeneracy and the Fermi sea contribution vanishes identically. Thermal disorder breaks inversion symmetry locally and lifts the Kramers degeneracy. In the present case however, the resulting contribution is entirely negligible. 

\section{Summary and Conclusions}
\label{Sec:S&C}

We have developed a method to calculate localized charge and spin currents in a multilayer system from the results of first-principles scattering calculations that include thermal lattice and spin disorder as well as chemical disorder for alloys. This allows us to factor out the effect of the interfaces that are unavoidable in scattering calculations and quantitatively evaluate parameters for bulk materials of interest in spin transport studies. We illustrated it by calculating the spin-flip diffusion length for Py $l_{\rm sf}^{\rm Py}=2.8\pm0.1$~nm and Pt $l_{\rm sf}^{\rm Pt}=5.3\pm0.4$~nm, the bulk spin polarization $\beta=0.75\pm0.01$ for Py and the spin Hall angle $\Theta_{\rm sH}=4.5\pm1.0\%$ for Pt at room temperature. Here the uncertainties were identified by systematically examining the approximations that must necessarily be made in calculations with finite computational resources. 

A comparison of the calculated bulk transport parameters with experimental results was inconclusive because, we believe, experiment is not able to unambiguously identify the bulk transport regime in the case of $l_{\rm sf}$ and $\Theta_{\rm sH}$ for Pt and many reported results are dominated by interface effects. Although recent attempts have been made to incorporate interface effects into the interpretation of experiments for bilayers, the effect of doing so appears to lead to diverging results rather than convergence \cite{Rojas-Sanchez:prl14, ZhangW:natp15, Nguyen:prl16, Tao:sca18, Berger:prb18b}.

The study presented in this paper opens up a wide range of possibilities to predict systematic trends for material parameters essential for spintronics applications. One possibility is to extend the calculations presented here to determine $l_{\rm sf}$ and $\beta$ for other bulk magnetic systems; to determine $\Theta_{\rm sH}$ and $l_{\rm sf}$ for other bulk 5$d$, 4$d$ and 3$d$ metals and their alloys all as a function of temperature with a view to identifying suitable candidates for spintronics applications and to better understand their temperature dependence and underlying scattering mechanisms.  Another very promising direction would be to use the localized spin currents to focus on interface effects and help disentangle bulk and interface contributions in the experimental studies we discussed briefly in the previous section.

\acknowledgements{K.G. is grateful to Yi Liu for help in starting this work and for supplying the Pt potentials with SOC included self consistently. This work was financially supported by the ``Nederlandse Organisatie voor Wetenschappelijk Onderzoek'' (NWO) through the research programme of the former ``Stichting voor Fundamenteel Onderzoek der Materie,'' (NWO-I, formerly FOM) and through the use of supercomputer facilities of NWO ``Exacte Wetenschappen'' (Physical Sciences). K.G. acknowledges funding from the Shell-NWO/FOM “Computational Sciences for Energy Research” PhD program (CSER-PhD; nr. i32; project number 13CSER059). The work was also supported by the Royal Netherlands Academy of Arts and Sciences (KNAW).}


\begin{thebibliography}{86}
\bibitem [{foo({\natexlab{a}})}]{footnote2}%
	\BibitemOpen
	\href@noop {} \bibinfo {note} {{In the framework of
			density functional theory \cite{Hohenberg:pr64, Kohn:pr65} these are the
			Kohn-Sham equations.}}\BibitemShut {Stop}%
\bibitem [{foo({\natexlab{b}})}]{footnote3}%
	\BibitemOpen
	\href@noop {} \bibinfo {note} {{If spin-orbit coupling
			is neglected, only the component of spin perpendicular to the magnetization
			is not conserved.}}\BibitemShut {Stop}%
\end{thebibliography}

%

\end{document}